\documentclass[aps,pre,twocolumn,showpacs,10pt]{revtex4-1}
\usepackage{color}
\usepackage{amsmath}
\usepackage{amsfonts}
\usepackage{amssymb}
\usepackage{graphicx}
\usepackage{float}
\usepackage{textcomp}

\begin{document}
\title{Reversible and Irreversible Aggregation of Magnetic Liposomes}
\author{Sonia Garc{\'i}a-Jimeno$^{1}$}
\author{Joan Estelrich$^1$}
\author{Jos{\'e} Callejas-Fern{\'a}ndez$^2$}
\author{S{\'a}ndalo Rold{\'a}n-Vargas$^{3,}$}
\email{sandalo@pks.mpg.de} 

\affiliation{$^1$ Secci{\'o} de Fisicoqu{\'i}mica, Facultat de Farm\`{a}cia i Ci\`{e}ncies de l'Alimentaci{\'o}, Universitat de Barcelona, Avda. Joan XXIII 17-31, E-08028, Barcelona, Catalonia, Spain,\\
$^2$ Grupo de F{\'i}sica de Fluidos y Biocoloides, Departamento de F{\'i}sica Aplicada, Facultad de Ciencias, Universidad de Granada, E-18071, Granada, Spain,\\
$^3$ Max Planck Institute for the Physics of Complex Systems,  D-01307, Dresden, Germany}

\begin{abstract}

Understanding stabilization and aggregation in magnetic nanoparticle systems is crucial to optimizing the functionality of these systems in real physiological applications. Here we address this problem for a specific, yet representative, system. We present an experimental and analytical study on the aggregation of superparamagnetic liposomes in suspension in the presence of a controllable external magnetic field. We study the aggregation kinetics and report an intermediate time power law evolution and a long time stationary value for the average aggregate diffusion coefficient, both depending on the magnetic field intensity. We then show that the long time aggregate structure is fractal with a fractal dimension that decreases upon increasing the magnetic field intensity. By scaling arguments we also establish an analytical relation between the aggregate fractal dimension and the power law exponent controlling the aggregation kinetics. This relation  is indeed independent on the magnetic field intensity. Despite the superparamagnetic character of our particles, we further prove the existence of a population of surviving aggregates able to maintain their integrity after switching off the external magnetic field. Finally, we suggest a schematic interaction scenario to rationalize the observed coexistence between reversible and irreversible aggregation.         
        
\end{abstract}

\maketitle

\section{INTRODUCTION}

In recent times, the \textit{ad hoc} design of novel mesoscopic particles has opened new research avenues and brought several promising applications. Significant examples of this bottom-up design appear in material science~\cite{Nie_general,Granick_book,dnapatchy1}, biotechnology~\cite{Condon_general,SEEMAN_03}, and nanomedicine~\cite{Wilczewska_general,Tran_general}. Indeed, the highly versatile functionality of these new primary constituents relies on our efficacy to control the distinct interactions governing their dynamic and structural properties.\\

A notable family among these new primary components is that constituted by those nano- and meso-sized particles able to respond to an external magnetic field. These ``magnetic nanodevices'' are usually categorized according to their remanent magnetization at a given temperature after having been exposed to an external magnetic field~\cite{Brown_ferro_super}. Thus mesoscopic particles consisting of single magnetic domains~\cite{Brown_book} can behave as permanent magnets due to their remanent (or even spontaneous) magnetization in the absence of an external magnetic field. This phenomenon is known as \textit{stable ferromagnetism}~\cite{Brown_ferro_super}. However, if thermal energy is able to cause the random orientation of the different single magnetic domains, the particle remanent magnetization after removing the external magnetic field will be negligible. These particles, which present no magnetic hysteresis, are known as \textit{superparamagnetic particles}~\cite{Brown_ferro_super,grains_1}. These two behaviors (ferromagnetic and superparamagnetic) are nowadays exploited in several consolidated research lines with a special emphasis in nanomedical applications~\cite{Reiss_magnetic_general,Pankhurst_magnetic_general,Tartaj_magnetic_general,Roca_magnetic_general,Arruebo_magnetic_general,Sun_magnetic_general,Mahmoudi_magnetic_general,Chomoucka_magnetic_general,Hu_magnetic_general}.\\

Among the distinct mesoscopic particles, liposomes (\textit{i.e.} mesosized lipid vesicles) have been recognized by their singular capabilities (\textit{e.g.} as drug delivery particles) due to their synthetically controllable size, surface electric charge, membrane elastic properties, and encapsulation efficiency~\cite{Lasic_1,Lasic_2,Lasic_3}. The superparamagnetic version of these highly tuneable particles results from our ability to encapsulate in their interior small (single domain) magnetite grains~\cite{Cuyper_1_sintesis,Martina_sintesis,Sabate_sintesis,Plassat_sintesis,Gomes_sintesis,Chen_sintesis,Chen_sintesis,Nappini_sintesis,Sonia_sintesis}. These are the so-called \textit{magnetic liposomes}. Thus these systems combine biocompatibility and vesicular structure~\cite{Barenholz} with their superparamagnetic character, therefore being magnetically controllable agents with no side-effects on the organism. Fruitful applications using these systems are already amenable to experimentation covering specific areas in therapy and diagnostics such as chemotherapy~\cite{Mahmoudi_magnetic_general,Hervault_chemotherapy}, hyperthermia~\cite{Safarikova_hyperthermia,Hamaguchi_hyperthermia,Ito_hyperthermia,Gonzales_hyperthermia,Bealle_hyperthermia}, magnetic resonance imaging~\cite{Bulte_resonance_imaging,Martina_sintesis,Du_resonance_imaging}, magnetic cell targeting~\cite{Dandamundi_magnetic_targeting,Martina_magnetic_targeting,Soenen_magnetic_targeting}, or magnetically driven delivery~\cite{Benyettou_delivery,Sonia_sintesis}.\\

Reaching an efficient functionality for these vesicular systems (and other magnetic nanoparticles) depends on our understanding of the distinct particle interactions. This understanding is intrinsically connected with those mechanisms controlling stabilization and aggregation. Indeed, magnetically induced aggregation not only provides us  with an implicit understanding on the particle interaction but it is explicitly manifested in real applications. For instance, aggregates of superparamagnetic particles present an enhanced heating efficiency in hyperthermia as compared to that corresponding to non-aggregated samples~\cite{Saville_2013,Saville_2014,Myrovali}.  The presence of aggregates can also increase the sensitivity of some detection techniques such as Surface-Enhanced Raman Scattering (SERS), leading to a significant increase of the Raman intensity~\cite{Jun}. Irreversible aggregates can also influence the system  functionality when their size is comparable to those length scales defining the targeted microenvironment, \textit{e.g.} in enhanced permeability and retention (EPR)~\cite{Matsumura,Arruebo_magnetic_general} or in the subsequent particle excretion from the body~\cite{Application_aggregation_1}. Having in mind this motivation, the main purpose of this work is to investigate the still poorly understood mechanisms controlling aggregation for a representative system of magnetic vesicles.\\

So far, magnetically induced aggregation has been experimentally investigated by different techniques being particularly focused on the study of superparamagnetic polystyrene particles. For instance, light scattering has been used to probe aggregation kinetics and/or aggregate structure~\cite{Licinio,Fernando_CSIA_2005,Fernando_JCP_2006,Fernando_PRE_2007,Fernando_JCIS_2008,Fernando_Langmuir_2009,Dominguez_2_morphology} whereas two-dimensional microscopy images have been analyzed to look into the cluster morphology~\cite{Fernando_CSIA_2005,Fernando_JCP_2006,Fernando_PRE_2008,Fernando_Langmuir_2009,Dominguez_1_morphology,Dominguez_2_morphology}. Apart from studies on real systems, simulations and analytical approaches have also been proposed to rationalize the aggregation of ferro- and superpara-magnetic particles where magnetic interaction is treated in terms of a dipolar hard-sphere like model~\cite{Faraudo_softmatter,Bertoni_simulation,Lorenzo_rings,Tlusty_Safran,Cerda_1_simulation}.\\

As far as we know, in this work we present the first comprehensive study on the aggregation of magnetic liposomes in suspension for a controllable external magnetic field. By Dynamic Light Scattering (DLS) we explore the aggregation kinetics and find an intermediate time regime where the aggregate diffusion coefficient presents a power law evolution. This evolution, which can be controlled by changing the magnetic field intensity, reaches at sufficiently long times a stationary value as a result of a competition between cluster formation and fragmentation. This final steady-state of the diffusion coefficient allows us to investigate the aggregate structure by Static Light Scattering (SLS). This structure is fractal and results in increasingly linear aggregates upon increasing the magnetic field intensity. Interestingly, we can appeal to scaling arguments and establish a relation between the aggregate fractal dimension and the power law exponent governing the aggregation kinetics. With this analytical approach we create a link between structure and dynamics in our system. To extend the aggregate characterization, we directly observe the system by Transmission Electron Microscopy (TEM) and report a coexistence between reversible and irreversible aggregates (\textit{i.e.} aggregates that survive despite switching off the external magnetic field). Finally, this coexistence is  discussed in terms of an interplay between interactions of different origin. Our results and the picture we offer may be of particular interest for predicting and controlling those time and length scales that play a relevant role in real physiological applications.\\  

The rest of the paper is organized as follows: In section II we introduce the system and present our methodologies. In section III we show and discuss our results on stabilization, aggregation kinetics, aggregate structure, and aggregate reversibility. Finally, in section IV we summarize our main findings and present our conclusions. 
 
\section{Materials and Methods}

The protocol for synthesizing the magnetic liposomes used in this work as well as part of the liposome characterization have been presented in a previous study ~\cite{Sonia_sintesis}. Here we summarize the previous methodologies and include the protocol to obtain our dynamic and static light scattering results as well as the methodology to acquire the TEM micrographs. Additional information on the characterization of the magnetic liposomes such as their zeta-potential at different salt concentrations or their encapsulation efficiency can be consulted in Ref.~\cite{Sonia_sintesis}. 

\subsection{Synthesis of Magnetic Liposomes}

\subsubsection{Materials}

Liposome membranes are constituted by Soybean phosphatidylcholine (PC) (Lipoid S-100), a zwitterionic phospholipid which was donated by Lipoid (Ludwigshafen, EU), and cholesterol (CHOL), which was purchased from Sigma (St. Louis, MO, USA). Nanoparticles of magnetite, stabilized with anionic coating (EMG 707), were purchased from FerroTec (Bedford, NH, USA) and have a nominal diameter of $10$ nm (determined by TEM),  a viscosity coefficient of less than  $5$ mPa$\cdot$s at $27$\textdegree{}C, and a $1.8$\% volume content of magnetite. 

\subsubsection{Preparation of Magnetic Liposomes}

Magnetic liposomes are obtained by using a modified version of the phase-reverse method~\cite{Szoka-Papahadjopoulos}. Lipids ($100$ $\mu$mols of PC and CHOL at $80:20$ molar ratio) dissolved in chloroform/methanol ($2:1$, v/v) are placed in a round-bottom flask and dried in a rotary evaporator under reduced pressure at $40$\textdegree{}C to form a thin film on the inner surface of the flask. The film is hydrated with $9$ ml of diethyl ether and $3$ ml of an aqueous dilution of FerroTec, resulting in a final concentration of $1.86$ g/l of magnetite. The mixture is then sonicated for $5$ min in a bath sonicator (Transsonic Digital Bath sonifier, Elma, EU) at $0$\textdegree{}C. Once the emulsion has been formed, it is placed in a round-bottom flask and the organic solution is removed under a pressure range of $420$-$440$ mmHg at room temperature. The emulsion becomes a gel and, finally, this gel transforms into a suspension of liposomes. Once the liposome suspension is obtained we add  $1$ ml of water, rotating the suspension at $760$ mmHg to remove the ether. Liposomes are then diluted with water until obtaining a final PC concentration of $16$ mmol/l. Liposome are then extruded both ways at room temperature into a Liposofast device (Avestin, Canada) through two polycarbonate membrane filters of $200$ nm pore size and for at least $9$ times ~\cite{MacDonald}. Separation of non-encapsulated ferrofluid from magnetic liposomes is performed by size exclusion chromatography (Sephacryl S-400 HR, GE Healthcare, Uppsala, Sweden). The iron content of the purified magnetic liposomes is determined by atomic absorption spectrophotometry (UNICAM PU 939 flame absorption spectrometer) giving an average value of $180$ $\mu$g/ml. PC was determined by colorimetry~\cite{Steward-Marshall}. Both determinations allow obtaining the $Fe^{3+}$/PC ratio which resulted in an average value of $45$ g/mol.

\subsection{Transmission Electron Microscopy}

To obtain the TEM micrographs of section III. E, we placed a drop of an initially stable aqueous suspension of magnetic liposomes at room temperature on a microscope slide covered with parafilm. The water  used  for  sample dilution was purified by inverse osmosis using Millipore equipment. To induce aggregation we placed at both sides of the microscope slide two Neodymium-Iron-Boron ($Nd_2Fe_{12} B$) magnets (Halde GAC, Barcelona, Spain). The magnetic field intensity created by the magnets in the space where the sample was placed is  $B = 80$ mT. After $15$ min of exposure to the magnetic field, a 400-mesh copper grid coated with a carbon film with a Formvar membrane was placed on the sample for $5$ min. After this time ($20$ min in total), the magnets were removed and a drop of water was added to the grid for washing the sample. This washing step was repeated once more. Then a drop containing a 2\% of uranyl acetate was added and, after $1$ min, the excess of staining solution was removed. The sample was allowed to dry in air for several minutes before observation. The observation was performed by transmission electron microscopy using an EFTEM (EM902 Zeiss, Carl Zeiss Jena, Germany) operating at $10^5$ V. In addition, TEM micrographs for non-aggregated samples were routinely used within this work for control purposes by using a transmission electron microscope Jeol 1010 (Jeol, Japan) operating at $8\cdot10^4$ V, recording the images by a Megaview III camera. The acquisition was accomplished with Soft-Imaging software (SIS, Germany).

\subsection{Magnetization of Magnetic Liposomes}

Magnetization curves of purified aqueous suspensions of magnetic liposomes as a function of the applied external magnetic field were obtained in a SQUID Quantum Design MPMS XL magnetometer. The probed external magnetic field ranged from $-600$ mT to $+600$ mT. Measurements were taken at room temperature.  

\subsection{Light Scattering Experiments}

The protocol we use to probe aggregation kinetics and aggregate structure under the influence of an external magnetic field by light scattering is partially similar to that reported in Refs.~\cite{Fernando_PRE_2007,Fernando_JCIS_2008,Fernando_Langmuir_2009} to study the aggregation of magnetic polystyrene particles. Here we present separately the experimental protocol to perform our measurements and a succinct theoretical background to interpret our measurements in terms of appropriate dynamic and static observables.      

\subsubsection{Experimental set-up and Measurement}

Light scattering experiments were performed by using a slightly modified Malvern 4700 System (UK), working with a He-Ne laser beam of  wavelength $\lambda = 632.8$ nm. To follow the dynamics of both aggregating and non-aggregating samples we perform DLS experiments at a fixed detection angle, $\theta_f = \pi/2$,  computing the scattered intensity autocorrelation function, $\left\langle I(\theta_f;t)I(\theta_f;t+\tau) \right\rangle$, for time intervals of $25$ s. Structure in our system is probed by SLS experiments which are performed by sweeping an angular detection range,  $\left[\theta_{min},\theta_{max} \right]$, by means of a movable photomultiplier arm where the average time scattered light intensity, $\left\langle I(\theta;t) \right\rangle$, is collected.\\   

For both aggregating and non-aggregating samples we used purified aqueous suspensions of magnetic liposomes where the presence of salt in the medium was prevented by inverse osmosis using Millipore equipment. We prepared sufficiently diluted suspensions at $0.1\%$ liposome volume fraction. This concentration avoids the effect of long-range interactions between liposomes in case of non-aggregating samples (section III.A) and gives us an optimal aggregation time for the magnetically induced aggregating samples. This time is sufficiently long compared with that needed for computing $\left\langle I(\theta_f;t)I(\theta_f;t+\tau) \right\rangle$ (2 orders of magnitude greater) but sufficiently short to follow the complete aggregation process. We ensured statistical reliability by performing at least $10$ independent experimental realizations of each DLS and SLS measurement for both aggregating and non-aggregating samples. In all the light scattering experiments temperature was kept constant at $25$\textdegree{}C.\\   

\begin{figure}[tb]
\center
\includegraphics[width=0.9\linewidth]{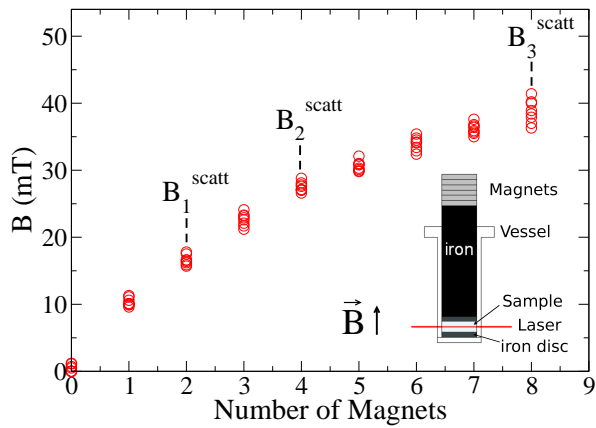} 
\caption{Magnetic field intensity, $B$, as a function of the number of Neodymium ($Nd_2Fe_{12}B$) magnets. Dashed lines signal the number of magnets (and therefore the magnetic field intensity, $B_i^{scatt}$, $i\in\lbrace 1,2,3\rbrace$) at which light scattering experiments were performed (sections III. B and C). Inset: Sketch of the experimental set-up.} 
\center
\label{setup} 
\end{figure}

The experimental set-up to induce liposome aggregation by means of an external magnetic field deserves further explanation. Figure~\ref{setup} shows a schematic view of this experimental set-up where the magnetic field intensity is controlled by adding or removing $Nd_2Fe_{12}B$ magnets on the top of the scattering vessel containing the sample. To enhance the magnetic field intensity acting on the sample, we insert between the pile of magnets and the sample a cylindrical iron bar to promote magnetic field line confinement. Thus, the direction of the magnetic field is essentially perpendicular to the scattering plane. Fig.~\ref{setup} also shows the magnetic field intensity acting on the sample as a function of the number of Neodymium magnets. We see how upon increasing the number of magnets the magnetic field intensity increases, leading to an intensity field saturation which imposes 
an upper threshold for the magnetic field intensity of about $40$ mT. Accordingly we performed DLS and SLS experiments for magnetically induced aggregating samples at $B^{scatt} = 16.6 (\pm 0.7)$, $27.5 (\pm 0.7)$, and $38.8 (\pm 0.6)$ mT. No aggregation was detected for $B < 16.6$ mT.

\subsubsection{Theoretical Background}

For non-aggregating samples ($B = 0$) we obtain the experimental liposome form factor, $P(q)$, through a SLS measurement by~\cite{pusey_1989,pusey_2002}:\\

\begin{equation}
P(q) = \frac{\left\langle I_{B = 0}(q;t) \right\rangle}{\left\langle I_{B = 0}(q_{min};t) \right\rangle} 
\end{equation}
\\
\noindent
Where $I_{B = 0}(q;t)$ is the previously mentioned scattered light intensity at time $t$ but expressed in terms of the modulus of the corresponding Fourier scattering vector $q = (4\pi n/\lambda)\sin(\theta/2)$ (where $q_{min}$ corresponds to $\theta_{min}$), being $n$ the refractive index of the scattering medium which here we take as $1.33$ (aqueous medium). For aggregating samples ($B \geq 16.6$ mT) we probe the structure of the magnetic liposome aggregates through their structure factor, $S(q)$~\cite{pusey_1989,pusey_2002,Fernando_JCP_2006,calcio,surface}:

\begin{equation}
S(q) = \frac{\left\langle I_{B \neq 0}(q;t) \right\rangle}{\left\langle I_{B = 0}(q;t) \right\rangle} 
\end{equation}
\\
\noindent
Where $I_{B \neq 0}(q;t)$ is the light intensity scattered by the aggregated sample at time $t$ for a given $q$ and for a magnetic field intensity $B \geq 16.6$ mT. We note that $I_{B = 0}(q;t)$ and $I_{B \neq 0}(q;t)$ correspond to the same sample before and after applying the magnetic field and, therefore, we should not introduce a relative density prefactor in Eq.(2)~\cite{pusey_1989,pusey_2002}. We also highlight that despite $I_{B \neq 0}(q;t)$ is measured in the presence of an external magnetic field, it presents a constant average value since our SLS measurements were performed once the samples had reached a stationary value for their average diffusion coefficient, therefore resulting in a non-evolving $S(q)$. This point is discussed in sections III.B and C.\\

Aggregates with fractal structure (section III.C) present a power law behavior for $S(q)$ within an intra-aggregate spatial scale which is constrained by the typical linear size of the aggregates and the linear size of the monomers (\textit{i.e.} the liposomes) constituting the aggregates~\cite{sorensen,calcio,surface}:

\begin{equation}
S(q) \sim q^{-d_f}  \,\,\ ;\,\,\ 1/R_{agg} \ll q \ll 1/\bar{a} 
\end{equation}
\\
\noindent
Where $d_f$ is the aggregate fractal dimension. Here $R_{agg}$ is the average aggregate radius whereas $\bar{a}$ is the average liposome radius. The $q$-range imposed by Eq.(3) results from the linear spatial dimensionality of $q^{-1}$ through the very definition of $q$ as a spatial frequency~\cite{sorensen}.\\

Dynamics in aggregating and non-aggregating samples is probed by DLS experiments through the intensity autocorrelation function $\left\langle I(q;t)I(q;t+\tau) \right\rangle$ at a fixed $q$. This autocorrelation function provides us with the corresponding electric field autocorrelation function, $g_E(\tau)$, by means of Siegert relation~\cite{berne_pecora}. In its turn, $g_E(\tau)$ is expanded into cumulants and interpreted in terms of a sample probability distribution of diffusion coefficients~\cite{berne_pecora,koppel}. The first cumulant, $\mu_1$, represents an inverse relaxation time containing both translational and rotational diffusive contributions~\cite{Dhont,Fernando_JCIS_2008}:

\begin{equation}
ln\left(g_E(\tau)\right)  = -\mu_1\tau + \mathcal{O}(\tau^2)  \,\,\ ;\,\,\ \mu_1 = D_tq^2 + 6D_r
\end{equation}
\\
\noindent
Where $D_t$ and $D_r$ are respectively the sample average translational and rotational diffusion coefficients, considered uncoupled by Eq. (4). We should also note that Eq.(4) assumes a simple exponential decay for $g_E(\tau)$ describing what would be in principle a probability distribution of relaxation times~\cite{stochastic} by a unique relaxation time, $1/\mu_1$. This simple exponential decay seems to be a good approximation for both aggregating and non-aggregating samples when treating the experimental $g_E(\tau)$. Moreover, in case of aggregating samples the typical aggregate size~\cite{Dhont} and the presence of an external magnetic field minimize the contribution of rotational diffusion~\cite{Fernando_JCIS_2008} in $g_E(\tau)$. For non-aggregating samples ($B = 0$), the spherical liposome shape directly excludes the presence of rotational diffusion in $g_E(\tau)$. Thus, in both cases, we assume:    

\begin{equation}
\mu_1 = D_{eff}^{\alpha}q^2  \,\,\ ;\,\,\ \alpha \in\lbrace B,0 \rbrace
\end{equation}
\\
\noindent
where $D_{eff}^B$ ($D_{eff}^0$) is the effective diffusion coefficient of an aggregating (non-aggretating) sample which essentially contains a translational contribution. More details on this approach as well as on more sophisticated treatments can be found in Refs.~\cite{Fernando_JCIS_2008,Fernando_JCP_2006}.

\section{Results}

\subsection{Characterization of Magnetic Liposomes}

\begin{figure}[tb]
\center
\hspace{0.2cm} \includegraphics[width=0.9\linewidth]{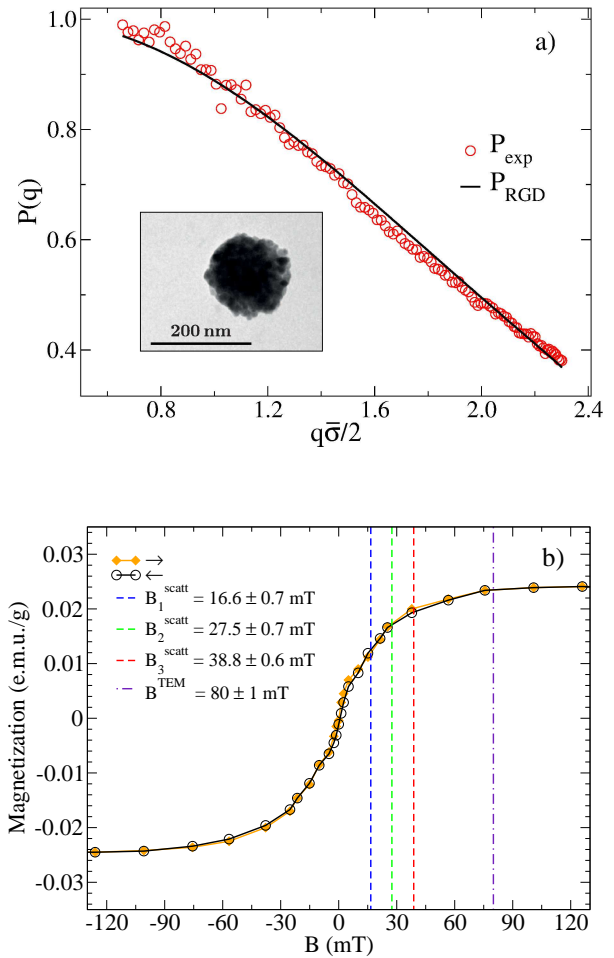}\vspace{1cm}
\includegraphics[width=0.9\linewidth]{Magnetization.eps} 
\caption{a) Form Factor, $P(q)$, of a diluted (non-aggregating) sample of magnetic liposomes. Circles stand for the experimental  values as obtained from a SLS measurement at $B = 0$ whereas solid line represents a fit according to a solid sphere RGD model and assuming a pseudo-Schulz distribution. Inset: TEM micrograph of a single magnetic liposome encapsulating a core of magnetite grains~\cite{Sonia_sintesis}. b) Magnetization as a function of the magnetic field intensity, $B$. Solid line with solid diamonds (empty circles) stands for the forward (backward) magnetization path. Both paths collapse into a single curve as a manifestation of no magnetic hysteresis. Dashed vertical lines signal the different field intensities ($B_i^{scatt}$, $i\in\lbrace 1,2,3\rbrace$, see also Fig.~\ref{setup}) at which light scattering experiments were performed (sections III. B and C) whereas the dotted-dashed vertical line indicates the high magnetic field intensity applied to the sample before obtaining the TEM micrographs of section III. E.} 
\center
\label{Characterization} 
\end{figure}

SLS measurements at $B = 0$ allow us to prove the stabilization of the non-aggregating samples and permit a characterization of the individual magnetic liposomes in terms of their shape, average size, and size polydispersity. In this respect, Figure~\ref{Characterization}a) shows the experimental form factor, $P(q)$, of a diluted sample of magnetic liposomes at $B = 0$ (Eq.(1)). The experimental $P(q)$ is here rationalized by means of a solid sphere model in the context of the Rayleigh-Gans-Debye (RGD) theory~\cite{Dhont} (solid line in Fig.~\ref{Characterization}a)). In particular, size polydispersity is introduced in the model by assuming a  three-modal distribution whose first five moments are distributed according to a Schulz distribution~\cite{calcio}. As a result, we obtain an average liposome diameter $\bar{\sigma} = 2\bar{a} = 180$ nm and a diameter polydispersity of $0.2$ (relative standard deviation divided by $\bar{\sigma}$). This $\bar{\sigma}$ value is in agreement with that obtained from the same sample by DLS experiments, where $D_{eff}^0$ (Eq.(5)) is interpreted in terms of the Stokes-Einstein relation~\cite{Dhont}. Moreover, liposomes observed by TEM micrographs~\cite{Sonia_sintesis} (\textit{e.g.} inset in Fig.~\ref{Characterization}a)) seem to present by simple inspection a size which is, roughly speaking, compatible with the $\bar{\sigma}$ obtained by $P(q)$.\\ 

At this point, one might ask for the repulsive interactions which avoid aggregation at $B = 0$. In this respect, two main interactions for stabilizing these and other lipid vesicle suspensions have been presented in the literature: Coulombic and hydration repulsions. On one hand, Coulombic repulsion, which is the main ingredient for stabilization in DLVO theory~\cite{DLVO_1,DLVO_2}, seems to be present in our system despite the non-polar nature of PC as accounted for by the weak but still non-negligible liposome zeta-potential~\cite{Sonia_sintesis}. On the other hand, short-range repulsive hydration forces have been associated to these and other lipid vesicles leading to stabilization even when Coulombic repulsion is not present~\cite{hydration_PC,Israelachvili_Wennerstrom,Petsev_Vekilov,Marcella,Ohki_Arnold,LeNeveu,surface}. Nevertheless, we should stress that the repulsive interactions stabilizing the system in the absence of an external magnetic field do not create long-range structural correlations between the liposomes for the probed dilution as manifested through $P(q)$ (which only contains correlations at the single particle level).\\         

To place the magnetic field intensities at which we perform our light scattering experiments and obtain our TEM micrographs for the aggregating samples, we present in Figure~\ref{Characterization}b) the magnetization, $M$, of the magnetic liposomes as a function of $B$ (see section II.C). A previous characterization of the liposome magnetization was already presented in Ref.~\cite{Sonia_sintesis}. We see how the forward and backward magnetization paths essentially collapse into a single curve: the magnetic liposomes do not present hysteresis. The absence of hysteresis represents a manifestation of the superparamagnetic nature of the magnetic liposomes which are indeed lipid vesicles encapsulating single-domain magnetite grains (linear size $\cong 10$ nm) which recover their random field orientation as soon as the magnetic field is switched off~\cite{grains_1}. As also shown in Fig.~\ref{Characterization}b), magnetization saturates around $\pm 100$ mT. In this respect, we see how our light scattering experiments (sections III.B and C) are performed below the saturation threshold whereas the TEM micrographs obtained for the aggregating samples (section III.E) correspond to an almost magnetically saturated sample. We also note that the different magnetic fields at which we perform our experiments for the aggregating samples do not present a significant difference in magnetization. However, the potential magnetic energy between magnetic liposomes could be significantly different for the different magnetic fields shown in Fig.~\ref{Characterization}b). In general, the potential magnetic energy between two magnetic particles (here liposomes) depends on the product of the dipole magnetic moments of the two particles, where each dipole magnetic moment is proportional to the particle magnetization~\cite{magnetic_interaction_1,magnetic_interaction_2,magnetic_interaction_3,chikazumi}. Therefore a given ratio between two different generic magnetizations, $M_1/M_2$, in Fig.~\ref{Characterization}b) will in general re-scale the potential magnetic energy between two magnetic particles by a factor $(M_1/M_2)^2$.\\     

\subsection{Aggregation Kinetics}

In this section we discuss the liposome aggregation dynamics under the influence of an external magnetic field by DLS measurements. Contrary to previous works on the aggregation of magnetic polystyrene particles~\cite{Fernando_CSIA_2005,Fernando_JCIS_2008,Fernando_JCP_2006}, aggregation is here induced by the external magnetic field with no added electrolyte. This is possible due to the weak Coulombic repulsive interaction between magnetic liposomes (see section III.A).\\

\begin{figure}[tb]
\center
\includegraphics[width=0.95\linewidth]{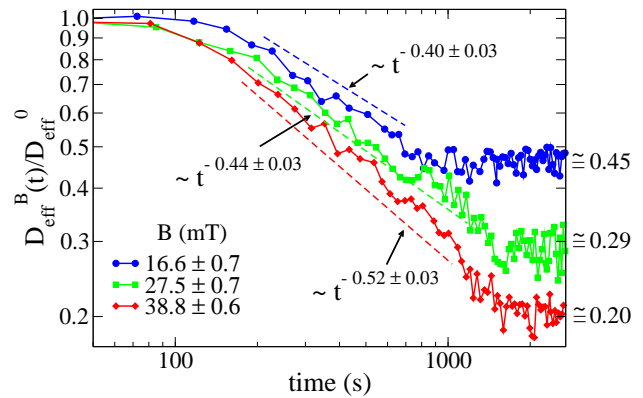} 
\caption{Log-log plot of the normalized diffusion coefficient, $D_{eff}^B/D_{eff}^0$, of an aggregating sample as a function of time for different magnetic field intensities (solid lines with solid symbols). Dashed lines represent the power law behavior at intermediate times, and for the different magnetic field intensities, before reaching a final stationary diffusion coefficient, $D_{eff}^B(t_{long})$. $D_{eff}^0$ is the single liposome diffusion coefficient as obtained from a (diluted) non-aggregating sample.} 
\center
\label{Diffusion} 
\end{figure}

Figure~\ref{Diffusion} shows the time evolution of $D_{eff}^B(t)$ at different magnetic field intensities (Eq.(5)). At short times $D_{eff}^B(t_{short}) \cong D_{eff}^0$: our aggregation process starts from a monomeric initial condition. At intermediate times $D_{eff}^B(t) \sim t^{-\alpha}$, where $\alpha$ is a $B$-dependent kinetic exponent which increases upon increasing magnetic field intensity. Finally, at long times $D_{eff}^B(t)$ reaches a plateau with a final stationary value, $D_{eff}^B(t_{long})$, which decreases upon increasing magnetic field intensity (from $D_{eff}^B(t_{long})/D_{eff}^0 \cong 0.45$ at $B = 16.6$ mT to $D_{eff}^B(t_{long})/D_{eff}^0 \cong 0.2$ at $B = 38.8$ mT).\\

The intermediate time power law behavior $D_{eff}^B(t) \sim t^{-\alpha}$ is a common feature in aggregation of mesoscopic particles which has been rationalized by different analytical approaches~\cite{leyvraz,kolb} being usually expressed in terms of the average aggregate size evolution $R_{agg}(t) \sim t^{\alpha}$. Depending on the system, this power law evolution will in principle continue without reaching a final stationary value~\cite{weitz_huang,lin,calcio} or it will present (like in our system) a final constant value for $D_{eff}^B(t)$ (or $R_{agg}(t)$) at sufficiently long times~\cite{Fernando_JCIS_2008,Fernando_JCP_2006,Faraudo_softmatter}. This second case has in general been interpreted as a balance between aggregation and fragmentation where the sample reaches a steady-state for the cluster-size distribution~\cite{Dongen_Ernst_Fragmentation,Fragmentation_Family}.\\

Balance between aggregation and fragmentation in magnetically induced aggregation processes has been discussed in terms of the so-called \textit{magnetic coupling parameter}, $\Gamma$, defined as the ratio (\textit{competition})
between magnetic dipole-dipole potential energy (which helps to retain particle bonds) and thermal energy (which tends to break particle bonds)~\cite{Faraudo_softmatter}:

\begin{equation}
\Gamma \equiv \frac{\mu m^2}{2\pi \bar{\sigma}^3k_BT} 
\end{equation}
\\
\noindent
Where $\mu$ is the medium magnetic permeability, $m$ the magnetic dipole moment of the particles, $T$ the absolute temperature, and $k_B$ the Boltzmann constant. By considering a proportionality between particle magnetization (Fig.~\ref{Characterization}b)) and particle magnetic dipole moment ~\cite{magnetic_interaction_3,chikazumi}, \textit{i.e.} $M \sim m$, we obtain $\Gamma \sim M^2$. This last relation leads us to an interesting result for understanding the $B$-dependence of $D_{eff}^B(t_{long})$. When comparing  in our system two different magnetizations (associated to two different magnetic field intensities, Fig.~\ref{Characterization}b)), with their corresponding  $D_{eff}^B(t_{long})$ we find:

\begin{equation}
\frac{\Gamma(B_i^{scatt})}{\Gamma(B_j^{scatt})} = \frac{M(B_i^{scatt})^2}{M(B_j^{scatt})^2} \cong \frac{D_{eff}^{B_j^{scatt}}(t_{long})}{D_{eff}^{B_i^{scatt}}(t_{long})} \,\,\ ;\,\ \forall i,j
\end{equation}
\\
\noindent
Thus, at constant temperature, re-scaling particle magnetization by a factor $\gamma$ will re-scale the final stationary diffusion coefficient by a factor $1/\gamma^2$, therefore connecting an individual particle property, $M$, with the final aggregate stability given by $D_{eff}^B(t_{long})$. For instance, $ \left(M(B_3^{scatt})/M(B_1^{scatt})\right)^2  \cong (1.55)^2$ (Fig.~\ref{Characterization}b)) to be compared with $D_{eff}^{B_1^{scatt}}(t_{long})/D_{eff}^{B_3^{scatt}}(t_{long}) = 2.25$ (Fig.~\ref{Diffusion}). This result, however, is satisfied by our system due to the low particle concentration where the influence of the liposome packing fraction is negligible~\cite{Faraudo_softmatter}.\\

\begin{figure}[tb]
\center
\includegraphics[width=0.9\linewidth]{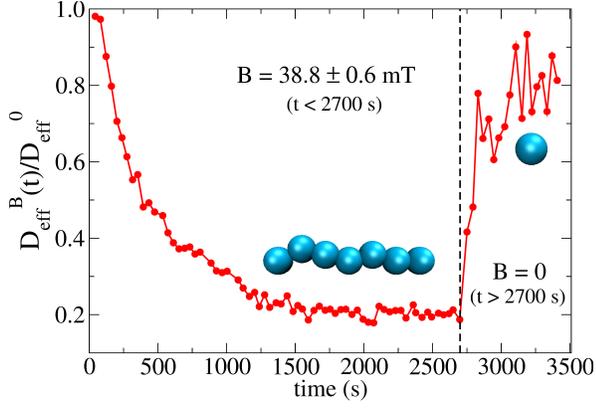} 
\caption{Linear-linear plot of the normalized diffusion coefficient, $D_{eff}^B/D_{eff}^0$, of an aggregating sample as a function of time. For times smaller than $2700$ s the sample aggregates under the influence of an applied field intensity $B = 38.8$ mT. For times greater than $2700$ s the external magnetic field is switched off and the sample almost recovers the single liposome diffusion coefficient, $D_{eff}^0$, as an indication of an almost complete disaggregation.} 
\center
\label{Break} 
\end{figure}  

To conclude this section we briefly anticipate the discussion on aggregation reversibility in our system when the magnetic field is switched off. Figure~\ref{Break} shows the time evolution of $D_{eff}^B(t)$ at $B = 38.8$ mT (\textit{i.e.} the highest field intensity in our DLS experiments) for times smaller than  2700 s and at $B = 0$ for times greater than 2700 s. Contrary to Fig.~\ref{Diffusion} where $B$ is maintained, Fig.~\ref{Break} shows how, as soon as the magnetic field is unplugged, $D_{eff}^{B=0}(t)$ tends to $D_{eff}^0$ as a manifestation of aggregation reversibility where the sample almost recovers its initial monomeric condition. However, for this magnetic field intensity, we cannot exclude by our DLS measurements and TEM micrographs  the presence of some small surviving aggregates after switching off the magnetic field (note that $D_{eff}^{B=0}(t) \lessapprox D_{eff}^0$). We will come back to this point in sections III.E and F.\\  
   
\subsection{Aggregate Structure}

We now proceed with the structural description of the liposome aggregates by SLS for those magnetic field intensities for which we already discussed the aggregation kinetics by DLS in the previous section. We stress that our SLS measurements are here performed in the presence of the magnetic field and for those (long) times at which the aggregate diffussion coefficient is already stabilized ($D_{eff}^B(t) = D_{eff}^B(t_{long})$). For this time regime we see no time evolution of the aggregate stucture factor, $S(q)$ (Eq.(2)).\\

\begin{figure}[tb]
\center
\includegraphics[width=0.9\linewidth]{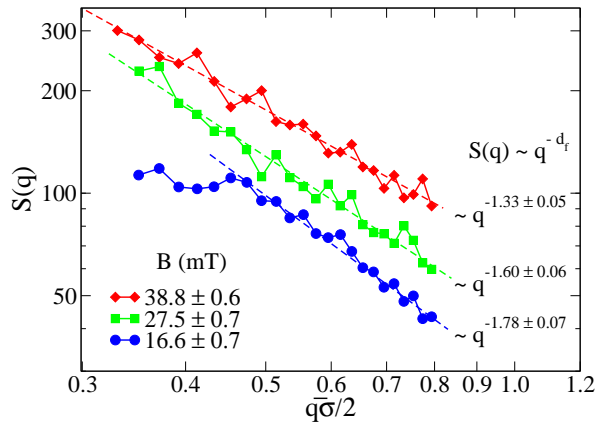} 
\caption{Log-log plot of the Structure Factor, $S(q)$, of an aggregated sample for different magnetic field intensities. Measurements are performed at long times, that is, when $D_{eff}^B(t) (= D_{eff}^B(t_{long}))$ is already stationary (see Fig.~\ref{Diffusion}). Dashed interpolation lines represent the expected fractal behavior, $S(q) \sim q^{-d_f},$ for the different field intensities.} 
\center
\label{Sq} 
\end{figure}

Figure~\ref{Sq} shows the structure factor, $S(q)$, for the aggregated samples at different magnetic field intensities. We see how the different $S(q)$'s present a power law fractal behavior within a certain intra-aggregate $q$-range according to Eq.(3). On one hand this range is right-side limited by the monomer (liposome) linear size, where $q < 2/\bar{\sigma}$. On the other hand we need an \textit{a priori} estimation for the left-side limit based on the linear size of the aggregates given by $R_{agg}$ (Eq.(3)). To estimate the left-side limit we consider Stokes-Einstein relation $R_{agg}/ \bar{\sigma} = (D_{eff}^B(t_{long})/ D_{eff}^0)^{-1}$, where here $R_{agg}$ is, formally speaking, the average hydrodynamic aggregate radius.\\

At low magnetic field intensities ($B = 16.6$ mT), the small aggregate linear size significantly restricts our $q$-range. Thus for $q\bar{\sigma}/2 \lesssim 0.5$ we already start to abandon the typical aggregate scale entering into the Guinier regime~\cite{sorensen}, therefore losing the details of the intra-aggregate structure whose spatial scale would be smaller than our $q^{-1}$ observational window. Although not reliable, the fractal dimension ($d_f \approxeq 1.78$) of the small aggregates at $B = 16.6$ mT would be compatible with that expected from a \textit{Diffusion Limited Cluster Aggregation} (DLCA)~\cite{lin}.\\   

Once we increase the magnetic field intensity the power law fractal behavior extends to smaller $q$ values due to the increasing aggregate size, therefore permitting a more reliable estimation of $d_f$. The effect is apparent: $d_f$ decreases upon increasing the magnetic field intensity. Thus, the increasing magnetic field intensity induces a highly directional magnetic liposome interaction which results in more linear fractal structures ($d_f \rightarrow 1$). In particular, we see how at $B = 38.8$ mT the resulting fractal dimension is $d_f = 1.33$. This value, which is far from the typical ramified aggregates reported in DLCA processes, is comparable with those obtained for magnetic polystyrene particles in the presence of a magnetic field with added electrolyte~\cite{Fernando_CSIA_2005,Fernando_JCP_2006} and compatible with simulations of dipolar hard-sphere fluids~\cite{Lorenzo_df,Camp_Patey}. Finally, we conclude this section by addressing the following question: how is the kinetic exponent, $\alpha$, connected with the aggregate fractal dimension, $d_f$?      

\subsection{Kinetic Exponent and Fractal Dimension}

Irreversible aggregation in diluted suspensions (like those studied here at intermediate times, Fig.~\ref{Diffusion}) can be understood in terms of a schematic binary reaction mechanism~\cite{brodie} of the form $A_i + A_j \rightarrow A_{i+j}$, where $A_i$ represents an aggregate constituted by $i$ monomers (here liposomes). In this context, scaling arguments can be applied to the rate coefficients, $k_{i,j}$, for the reaction between $A_i$ and $A_j$ aggregates~\cite{Dongen_Ernst_lambda_1,Dongen_Ernst_lambda_2}:

\begin{equation}
k_{ai,aj} = a^{\lambda} k_{i,j}  \,\,\ ;\,\,\ \forall i,j \,\,\ ;\,\,\ a \in \mathbb{N} \,\,\ ;\,\,\ \lambda \in \mathbb{R}
\end{equation}
\\
\noindent
Where we assume a homogeneous behavior for the rate coefficients, $k_{i,j}$, through a homogeneity parameter, $\lambda$, which is restricted by $\lambda \leqslant 1$ for non-gelling aggregation processes. For those processes  where reactions between small-small and large-large aggregates are equally probable (\textit{e.g.} DLCA) we have $\lambda = 0$, resulting in a balance between aggregate collision cross section (which increases upon increasing $i$) and aggregate diffusivity (which decreases upon increasing $i$). Those processes where $\lambda < 0$ ($\lambda > 0$) result in a more likely reaction between small-small (large-large) aggregates. Equation (8) implies a power law behavior for the average number of monomers  per aggregate at time $t$, $\bar{n}(t)$~\cite{Dongen_Ernst_lambda_1}:

\begin{equation}
\bar{n}(t) = \sum_{i=1} i{\cal N}_i(t) \sim t^{1/(1-\lambda)} \,\ ;\,\ \left( \sum_{i=1} {\cal N}_i(t) \equiv 1 \right)   
\end{equation}
\\
\noindent
Where ${\cal N}_i(t)$ is the relative frequency of aggregates constituted by $i$ monomers at time $t$. If we now incorporate the fractal scaling of the aggregates according to their fractal dimension, $\bar{n}(t) \sim R_{agg}(t)^{d_f}$ (previous section), and assume Stokes-Einstein relation, $R_{agg}(t) \sim D_{eff}^B(t)^{-1}$, we reach:

\begin{equation}
D_{eff}^B(t) \sim t^{-1/(1-\lambda)d_f} 
\end{equation}
\\
\noindent
Where we immediately recognize the intermediate time power law behavior discussed in section III.B (Fig.~\ref{Diffusion}) with $\alpha \equiv 1/(1-\lambda)d_f$.\\ 

\begin{table}[!htbp]
%\scriptsize
\caption{\textbf{Homogeneity parameter, $\lambda$, kinetic exponent, $\alpha$, and fractal dimension, $d_f$, for different magnetic field intensities.}} \,\,\,\,\,\
\begin{tabular}{|c|c|c|c|}
\hline      
          $B$ (mT)         & \,\,\   $16.6 \pm 0.7$   \,\,\ &  \,\,\   $27.5 \pm 0.7$   \,\,\ &  \,\,\   $38.8 \pm 0.6$   \,\,\ \\
\hline                    
    \,\,\ $\lambda$  \,\,\ &  \,\,\ $-0.40 \pm 0.12$  \,\,\ &  \,\,\ $-0.42 \pm 0.11$  \,\,\ &  \,\,\ $-0.45 \pm 0.10$  \,\,\ \\
\hline                    
    \,\,\ $\alpha$  \,\,\ &  \,\,\ $0.40 \pm 0.03$  \,\,\ &  \,\,\ $0.44 \pm 0.03$  \,\,\ &  \,\,\ $0.52 \pm 0.03$  \,\,\ \\
\hline                    
    \,\,\ $d_f$  \,\,\ &  \,\,\ $1.78 \pm 0.07$  \,\,\ &  \,\,\ $1.60 \pm 0.06$  \,\,\ &  \,\,\ $1.33 \pm 0.05$  \,\,\ \\    
\hline
%\multicolumn{}{p{} }{}
\end{tabular}
\label{tab_lambda}
\end{table}  

Table~\ref{tab_lambda} shows the homogeneity parameter, $\lambda$, for the different magnetic field intensities at which we previously discussed the aggregation kinetics and the aggregate structure (included in the table are $\alpha$, Fig.~\ref{Diffusion}, and $d_f$, Fig.~\ref{Sq}). It is interesting to note from the table how $\lambda$ is almost independent ($\lambda \cong -0.4$) on the applied magnetic field intensity. As a result, here all the aggregation processes at intermediate times present a connection between their corresponding $\alpha$ and $d_f$ values which leads to a common $\bar{n}(t)$ behavior given by Eq.(9). According to our previous discussion, we can reach a physical intuition for the negative $\lambda$ value by considering that the decreasing aggregate diffusivity is not \textit{compensated} by the increasing aggregate collision cross section as $\bar{n}(t)$ increases. Indeed, the very geometry of the magnetic field lines around an aggregate results in an almost constant (elongated) cross section which will not depend on $\bar{n}(t)$~\cite{Fernando_PRE_2008}. However, diffusivity will decrease upon increasing $\bar{n}(t)$ with an expected power law evolution~\cite{Fernando_PRE_2008}. These scaling behaviors therefore lead to a more efficient reaction between small-small aggregates as compared with that between large-large aggregates.  

\subsection{TEM Micrographs}

We now proceed to discussing the TEM micrographs obtained from the magnetic liposome suspensions after applying an intense magnetic field of $B = 80$ mT (see section II.B and Fig.~\ref{Characterization}b) in section III.A). This magnetic field intensity is much higher than that applied during our light scattering experiments and almost corresponds to the liposome magnetic saturation (Fig.~\ref{Characterization}b)). It is important to stress that before capturing our TEM micrographs, the sample first aggregates according to the protocol described in section II.B and then, after being exposed to the magnetic field, the magnetic field is removed leaving the sample to evolve for several minutes  (this time is indeed significantly greater than that needed to recover the almost monomeric state reported in Fig.~\ref{Break}). The discussion we present here is based on a simple observational inspection of the TEM micrographs.\\

\begin{figure}[!htbp]
\center
\includegraphics[width=0.9\linewidth]{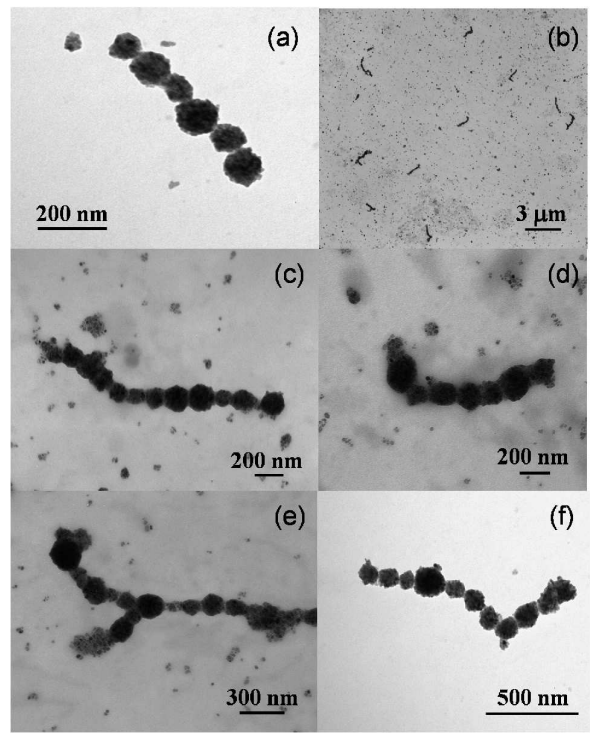} 
\caption{TEM micrographs of a sample of magnetic liposomes which was first exposed to an intense magnetic field of $B = 80$ mT. The sample then evolved for several minutes without the presence of an external magnetic field before capturing the images (see sections II.B and III.A, and  Fig.~\ref{Characterization}b)). The micrographs correspond to different control regions and different magnifications.} 
\center
\label{Pannel_TEM} 
\end{figure}

Figure~\ref{Pannel_TEM} shows different TEM micrographs of the liposome suspension for different control regions within the sample and different magnifications. The first message is obvious: despite having evolved without the presence of an external magnetic field, the sample shows the existence of several surviving aggregates (Fig.~\ref{Pannel_TEM}b)). From now on we will refer to these aggregates as \textit{irreversible aggregates}. Despite we cannot discard the existence of small surviving aggregates after applying lower magnetic field intensities, the presence of these irreversible aggregates seems to contrast with the almost complete reversible aggregation reported at the end of section III.B (Fig.~\ref{Break}). Micrographs also support a second \textit{structural} message: irreversible aggregates show an almost linear structure (Fig.~\ref{Pannel_TEM}a),c),d) and f)) compatible with the fractal dimension ($d_f \rightarrow 1$) that would be expected after having aggregated under the influence of an intense magnetic field (see section III.C). Indeed, the only non-linear (branched) structures we see (albeit scarce) correspond to "Y-like" shaped aggregates where one of the liposomes acts as a junction point between two branches (Fig.~\ref{Pannel_TEM}e))~\cite{Tlusty_Safran}.\\

We now discuss some specific but still significant details. On one hand, aggregate size polydispersity seems to be rather low (with an average number of liposomes per aggregate of the order of $10$). On the other hand, irreversible aggregates seem to be constituted by rather monodisperse liposomes, that is, the size polydispersity of the liposomes forming the irreversible aggregates seems to be lower than that corresponding to the whole sample (section III.A). This rather monodisperse aggregate composition is compatible with theoretical predictions for chain-like aggregates in polydisperse ferrofluids where the presence of small magnetic particles as part of the aggregates is not favorable~\cite{Kantorovich_1,Kantorovich_2}. On this theoretical basis, we could understand the presence of small magnetite spots in our TEM micrographs as a manifestation of small dried magnetic liposomes which were not able of being part of the irreversible aggregates.\\ 

Aggregate shape also deserves further discussion. Magnetic particles with remanent magnetization in the absence of an external magnetic field can in principle self-assemble into closed aggregates. In particular, computational studies on dipolar hard-spheres~\cite{Lorenzo_rings} and experimental investigations with microscopic ferromagnetic particles~\cite{Fernando_rings} show the emergence of ring shaped aggregates. However, our irreversible aggregates do not show (at least from the current TEM micrographs)  ring structures. The absence of rings (whose presence is expected for particles with a high remanent magnetization) can represent a manifestation of the superparamagnetic nature of the magnetic liposomes for which no magnetic hysteresis was detected (Fig.~\ref{Characterization}b)). In this respect, and giving that we cannot appeal to particle remanent magnetization, what is the interaction mechanism responsible for maintaining the integrity of our irreversible aggregates in the absence of an external magnetic field?                  

\subsection{Liposome Interactions}

Reaching a precise quantitative answer to the previous question needs further systematic investigation, specially focused on the empirical phenomenology associated to the different interactions governing aggregation and stabilization in our system. Instead, here we address this question by briefly discussing a plausible schematic picture based on purely heuristic arguments.\\

Coexistence between reversible and irreversible aggregation in mesoscopic particle systems has been rationalized by the existence of primary and secondary minima of the particle potential energy~\cite{primary_secondary_minima_1}. Thus, when aggregation is promoted by a certain mechanism (\textit{e.g.} here by applying an external magnetic field), particles can in principle aggregate in a permanent (irreversible) state which is associated to a primary minimum where the aggregated state will be maintained despite canceling the mechanism provoking aggregation (\textit{e.g.} by switching off the external magnetic field). However, particles can also aggregate in a secondary minimum being restored to their non-aggregated state as soon as the mechanism promoting aggregation is canceled.\\       

\begin{figure}[!htbp]
\center
\includegraphics[width=0.9\linewidth]{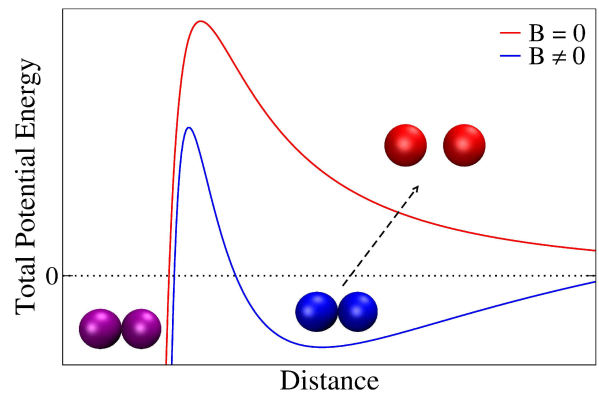} 
\caption{Sketch of the total potential energy between two magnetic particles (here liposomes) based on the theoretical approximation of Refs.~\cite{primary_secondary_minima_1,primary_secondary_minima_2,primary_secondary_minima_3}. Blue line represents the total potential energy in the presence of an external magnetic field where particles can become stuck in a primary (purple particles) or in a secondary (blue particles) minimum. In the absence of an external magnetic field (red line) those particles that were in a secondary minimum become separated (reversible aggregation, red particles) whereas particles that were in a primary minimum retain their aggregated state (irreversible aggregation, purple particles). Distance can here be interpreted as the separation distance between the external surface of the particles (\textit{i.e.} the distance between the external surface of two liposome membranes). Separation between particles sketches is merely illustrative in the figure: thus separation between non-aggregated particles (red) has been enhanced whereas that corresponding to the aggregated particles (blue and purple) has been intentionally reduced.} 
\center
\label{interactions} 
\end{figure}

The idea of an interaction mechanism based on the existence of primary and secondary minima to understand irreversible and reversible aggregation is schematically presented in Figure~\ref{interactions} for a magnetically induced aggregation process. Thus, in the presence of an external magnetic field (blue line) some particles (purple) aggregate in a permanent (irreversible) primary minimum whereas other (blue particles) aggregate in a (reversible) secondary minimum. When the external magnetic field is switched off (red line), particles aggregating in the secondary minimum become separated. Theoretical approximations based on this underlying picture have been proposed in the past to understand the aggregation of superparamagnetic colloidal latex particles~\cite{primary_secondary_minima_1,primary_secondary_minima_2,primary_secondary_minima_3}. In this context, the emergence of primary and secondary minima results from the interplay (or competition) between Coulombic repulsion (treated by a linear superposition approximation), London-van der Waals attraction (Derjaguin approach), and magnetic dipole-dipole attraction. This approach has indeed shown to be successful for predicting and controlling magnetic flocculation to concentrate or remove ultrafine magnetic particles (linear size smaller than $5$ $\mu$m)~\cite{review_magnetic_flocculation}.\\   

These interactions~\cite{primary_secondary_minima_1,primary_secondary_minima_2,primary_secondary_minima_3} (\textit{i.e.} DLVO and magnetic dipole-dipole interactions) seem to play, \textit{a priori}, a significant role by governing stabilization-aggregation in our magnetic liposome system. Thus, Coulombic repulsion is present in our system as manifested by the non-negligible zeta-potential~\cite{Sonia_sintesis} whereas London-van der Waals interaction has been identified as the main short range attraction between lipid membranes~\cite{Israelachvili}. In addition, a non-DLVO ingredient widely reported in the liposome literature should presumably be considered to reach a complete theoretical description for the magnetic liposome aggregation mechanism:  short range hydration repulsion~\cite{hydration_PC,Israelachvili_Wennerstrom,Petsev_Vekilov,Marcella,Ohki_Arnold,LeNeveu,surface}.\\

To calibrate whether or not this \textit{complete} approach is consistent with a primary-secondary minimum scenario in the present system, additional experiments should be performed. In particular, a more refined control of the magnetic field intensity would help us to better quantify the emergence of primary and secondary minima. Moreover, further experiments in the presence of added electrolyte could also help us to judiciously manipulate Coulombic and hydration repulsions~\cite{surface}, therefore providing valuable quantitative information on the interplay between attractive and repulsive interactions. In the meantime, we are led to speculate on the primary-secondary minimum picture as a plausible mechanism to explain irreversibility-reversibility in our system suggesting future systematic experimental work to resolve this issue further.\\

\section{SUMMARY AND CONCLUSIONS}

We have presented a comprehensive study on the aggregation of superparamagnetic liposomes in solution under the influence of a controllable external magnetic field. We have investigated the liposome aggregation kinetics, the aggregate structure, and the coexistence between reversible and irreversible aggregation by Dynamic and Static Light Scattering (DLS and SLS), and by images obtained from Transmission Electron Microscopy (TEM).\\

Aggregation kinetics has been probed by DLS and followed by the time evolution of the aggregate diffusion coefficient. For a constant magnetic field intensity, the aggregate diffusion coefficient shows a stationary value at sufficiently long times which decreases upon increasing the external magnetic field intensity. We have proven how this stationary value, which is here interpreted as a balance between liposome cluster aggregation and fragmentation, scales with the square of the liposome magnetization. Before reaching its stationary value, the diffusion coefficient follows a time dependent power law behavior with a kinetic exponent, $\alpha$, which increases upon increasing magnetic field intensity. As a manifestation of aggregation reversibility, we have further shown how liposomes aggregating under the influence of a low magnetic field intensity ($< 40$ mT) almost recover their initial (non-aggregated) state when the external magnetic field is switched off.\\

We have taken advantage of the long time stationary value of the liposome aggregate diffusion coefficient to probe the aggregate structure by SLS through the aggregate structure factor. Thus we have proven the aggregate structure to be fractal and shown how the fractal dimension, $d_f$, decreases upon increasing the external magnetic field intensity, resulting in the emergence of almost linear aggregate structures ($d_f \rightarrow 1$). We have finally shown how structure and dynamics are connected in our system by finding a scaling relation between the kinetic exponent, $\alpha$, and the aggregate fractal dimension, $d_f$, which allows us to understand aggregation kinetics and aggregate structure in terms of a single homogeneity parameter.\\

By TEM micrographs we have also shown the existence of irreversible liposome aggregates which result from an aggregation process in the presence of an intense external magnetic field ($80$ mT). These irreversible aggregates show an open linear structure and survive despite switching off the external magnetic field. To rationalize the coexistence between reversible and irreversible aggregates, we have suggested a schematic picture based on the existence of primary and secondary minima of the liposome potential energy.\\

In conclusion, we have revealed the rich interaction scenario involved in the magnetically induced aggregation of superparamagnetic liposomes in suspension. Understanding the mechanisms controlling the aggregation of these (and other) biocompatible magnetic nanodevices is a cornerstone for exploiting their singular capabilities as functional agents in promising medical and biotechnological applications.  

\section{ACKNOWLEDGEMENTS}

We thank Miguel Hern{\'a}ndez-D{\'i}az, Fernando Vereda, Miguel P{\'e}laez-Fern{\'a}ndez, and Daniel Aguilar-Hidalgo for their valuable technical assistance. We also thank the scientific-technical services of the University of Granada and the University of Barcelona for their support and assistance with the TEM micrographs. Particle sketches in Fig.~\ref{Break} and Fig.~\ref{interactions} were made with VMD software support (VMD is developed with NIH support by the Theoretical and Computational Biophysics group at the Beckman Institute, University of Illinois at Urbana-Champaign). J.C.-F. acknowledges support from Ministerio de Econom{\'i}a y Competitividad (MINECO), Plan Nacional de Investigaci{\'o}n, Desarrollo e Innovaci{\'o}n Tecnol{\'o}gica (I + D + i), Project FIS2016-80087-C2-1-P. We would like to express our gratitude to Fernando Mart{\'i}nez-Pedrero, Lorenzo Rovigatti, Izaak Neri, and Jakob L\"{o}ber for the critical reading of this manuscript and their valuable comments.

\bibliographystyle{apsrev}

\begin{thebibliography}{0}
\expandafter\ifx\csname natexlab\endcsname\relax\def\natexlab#1{#1}\fi
\expandafter\ifx\csname bibnamefont\endcsname\relax
  \def\bibnamefont#1{#1}\fi
\expandafter\ifx\csname bibfnamefont\endcsname\relax
  \def\bibfnamefont#1{#1}\fi
\expandafter\ifx\csname citenamefont\endcsname\relax
  \def\citenamefont#1{#1}\fi
\expandafter\ifx\csname url\endcsname\relax
  \def\url#1{\texttt{#1}}\fi
\expandafter\ifx\csname urlprefix\endcsname\relax\def\urlprefix{URL }\fi
\providecommand{\bibinfo}[2]{#2}
\providecommand{\eprint}[2][]{\url{#2}}

\bibitem[{\citenamefont{Nie et~al.}(2010)\citenamefont{Nie, Petukhova, and
  Kumacheva}}]{Nie_general}
\bibinfo{author}{\bibfnamefont{Z.}~\bibnamefont{Nie}},
  \bibinfo{author}{\bibfnamefont{A.}~\bibnamefont{Petukhova}},
  \bibnamefont{and}
  \bibinfo{author}{\bibfnamefont{E.}~\bibnamefont{Kumacheva}},
  \bibinfo{journal}{Nat. Nano.} \textbf{\bibinfo{volume}{5}},
  \bibinfo{pages}{15} (\bibinfo{year}{2010}).

\bibitem[{Gra(2012)}]{Granick_book}
\emph{\bibinfo{title}{{Janus Particles Synthesis, Self-Assembly, and
  Applications, {\rm RSC Smart Materials, Edited by S. Granick and S. Jiang}}}}
  (\bibinfo{publisher}{RCS Publishing, London}, \bibinfo{year}{2012}).

\bibitem[{\citenamefont{Wang et~al.}(2012)\citenamefont{Wang, Wang, Breed,
  Manoharan, Feng, Hollingsworth, Weck, and Pine}}]{dnapatchy1}
\bibinfo{author}{\bibfnamefont{Y.}~\bibnamefont{Wang}},
  \bibinfo{author}{\bibfnamefont{Y.}~\bibnamefont{Wang}},
  \bibinfo{author}{\bibfnamefont{D.}~\bibnamefont{Breed}},
  \bibinfo{author}{\bibfnamefont{V.}~\bibnamefont{Manoharan}},
  \bibinfo{author}{\bibfnamefont{L.}~\bibnamefont{Feng}},
  \bibinfo{author}{\bibfnamefont{A.}~\bibnamefont{Hollingsworth}},
  \bibinfo{author}{\bibfnamefont{M.}~\bibnamefont{Weck}}, \bibnamefont{and}
  \bibinfo{author}{\bibfnamefont{D.}~\bibnamefont{Pine}},
  \bibinfo{journal}{Nature} \textbf{\bibinfo{volume}{491}}, \bibinfo{pages}{51}
  (\bibinfo{year}{2012}).

\bibitem[{\citenamefont{Condon}(2006)}]{Condon_general}
\bibinfo{author}{\bibfnamefont{A.}~\bibnamefont{Condon}},
  \bibinfo{journal}{Nat. Rev. Genet.} \textbf{\bibinfo{volume}{7}},
  \bibinfo{pages}{565} (\bibinfo{year}{2006}).

\bibitem[{\citenamefont{Seeman}(2003)}]{SEEMAN_03}
\bibinfo{author}{\bibfnamefont{N.}~\bibnamefont{Seeman}},
  \bibinfo{journal}{Nature} \textbf{\bibinfo{volume}{421}},
  \bibinfo{pages}{427} (\bibinfo{year}{2003}).

\bibitem[{\citenamefont{Wilczewska et~al.}(2012)\citenamefont{Wilczewska,
  Niemirowicz, Markiewicz, and Car}}]{Wilczewska_general}
\bibinfo{author}{\bibfnamefont{A.}~\bibnamefont{Wilczewska}},
  \bibinfo{author}{\bibfnamefont{K.}~\bibnamefont{Niemirowicz}},
  \bibinfo{author}{\bibfnamefont{K.}~\bibnamefont{Markiewicz}},
  \bibnamefont{and} \bibinfo{author}{\bibfnamefont{H.}~\bibnamefont{Car}},
  \bibinfo{journal}{Pharmacol. Rep.} \textbf{\bibinfo{volume}{64}},
  \bibinfo{pages}{1020} (\bibinfo{year}{2012}).

\bibitem[{\citenamefont{Tran and Webster}(2010)}]{Tran_general}
\bibinfo{author}{\bibfnamefont{N.}~\bibnamefont{Tran}} \bibnamefont{and}
  \bibinfo{author}{\bibfnamefont{T.}~\bibnamefont{Webster}},
  \bibinfo{journal}{J. Mater. Chem.} \textbf{\bibinfo{volume}{20}},
  \bibinfo{pages}{8760} (\bibinfo{year}{2010}).

\bibitem[{\citenamefont{Brown~Jr.}(1963)}]{Brown_ferro_super}
\bibinfo{author}{\bibfnamefont{W.}~\bibnamefont{Brown~Jr.}},
  \bibinfo{journal}{Phys. Rev.} \textbf{\bibinfo{volume}{130}},
  \bibinfo{pages}{1677} (\bibinfo{year}{1963}).

\bibitem[{\citenamefont{Brown~Jr.}(1962)}]{Brown_book}
\bibinfo{author}{\bibfnamefont{W.}~\bibnamefont{Brown~Jr.}},
  \emph{\bibinfo{title}{Magneto-static Principles in Ferronzagnetism, Chap. 6}}
  (\bibinfo{publisher}{North-Holland Publishing Company, Amsterdam},
  \bibinfo{year}{1962}).

\bibitem[{\citenamefont{Cabuil}(2000)}]{grains_1}
\bibinfo{author}{\bibfnamefont{V.}~\bibnamefont{Cabuil}},
  \emph{\bibinfo{title}{Preparation and Properties of Magnetic Nanoparticles}}
  (\bibinfo{publisher}{Encyclopedia of Surface and Colloid Science, Marcel
  Dekker}, \bibinfo{year}{2000}).

\bibitem[{\citenamefont{Reiss and Hutten}(2005)}]{Reiss_magnetic_general}
\bibinfo{author}{\bibfnamefont{G.}~\bibnamefont{Reiss}} \bibnamefont{and}
  \bibinfo{author}{\bibfnamefont{A.}~\bibnamefont{Hutten}},
  \bibinfo{journal}{Nat. Mater.} \textbf{\bibinfo{volume}{4}},
  \bibinfo{pages}{725} (\bibinfo{year}{2005}).

\bibitem[{\citenamefont{Pankhurst et~al.}(2003)\citenamefont{Pankhurst,
  Connolly, Jones, and Dobson}}]{Pankhurst_magnetic_general}
\bibinfo{author}{\bibfnamefont{Q.}~\bibnamefont{Pankhurst}},
  \bibinfo{author}{\bibfnamefont{J.}~\bibnamefont{Connolly}},
  \bibinfo{author}{\bibfnamefont{S.}~\bibnamefont{Jones}}, \bibnamefont{and}
  \bibinfo{author}{\bibfnamefont{J.}~\bibnamefont{Dobson}},
  \bibinfo{journal}{J. Phys. D} \textbf{\bibinfo{volume}{36}},
  \bibinfo{pages}{R167} (\bibinfo{year}{2003}).

\bibitem[{\citenamefont{Tartaj et~al.}(2003)\citenamefont{Tartaj, del
  Puerto~Morales, Veintemillas-Verdaguer, and Serna}}]{Tartaj_magnetic_general}
\bibinfo{author}{\bibfnamefont{P.}~\bibnamefont{Tartaj}},
  \bibinfo{author}{\bibfnamefont{M.}~\bibnamefont{del Puerto~Morales}},
  \bibinfo{author}{\bibfnamefont{T.}~\bibnamefont{Veintemillas-Verdaguer},
  \bibfnamefont{S.~Gonz{\'a}lez-Carre{\~n}o}}, \bibnamefont{and}
  \bibinfo{author}{\bibfnamefont{C.}~\bibnamefont{Serna}}, \bibinfo{journal}{J.
  Phys. D} \textbf{\bibinfo{volume}{36}}, \bibinfo{pages}{R182}
  (\bibinfo{year}{2003}).

\bibitem[{\citenamefont{Roca et~al.}(2009)\citenamefont{Roca, Costo, Rebolledo,
  Veintemillas-Verdaguer, Tartaj, Gonz{\'a}lez-Carre{\~n}o, Morales, and
  Serna}}]{Roca_magnetic_general}
\bibinfo{author}{\bibfnamefont{A.}~\bibnamefont{Roca}},
  \bibinfo{author}{\bibfnamefont{R.}~\bibnamefont{Costo}},
  \bibinfo{author}{\bibfnamefont{A.~F.} \bibnamefont{Rebolledo}},
  \bibinfo{author}{\bibfnamefont{S.}~\bibnamefont{Veintemillas-Verdaguer}},
  \bibinfo{author}{\bibfnamefont{P.}~\bibnamefont{Tartaj}},
  \bibinfo{author}{\bibfnamefont{T.}~\bibnamefont{Gonz{\'a}lez-Carre{\~n}o}},
  \bibinfo{author}{\bibfnamefont{M.}~\bibnamefont{Morales}}, \bibnamefont{and}
  \bibinfo{author}{\bibfnamefont{C.}~\bibnamefont{Serna}}, \bibinfo{journal}{J.
  Phys. D: Appl. Phys.} \textbf{\bibinfo{volume}{42}}, \bibinfo{pages}{224002}
  (\bibinfo{year}{2009}).

\bibitem[{\citenamefont{Arruebo et~al.}(2007)\citenamefont{Arruebo,
  Fern{\'a}ndez-Pacheco, Ibarra, and
  Santamar\'{\i}a}}]{Arruebo_magnetic_general}
\bibinfo{author}{\bibfnamefont{M.}~\bibnamefont{Arruebo}},
  \bibinfo{author}{\bibfnamefont{R.}~\bibnamefont{Fern{\'a}ndez-Pacheco}},
  \bibinfo{author}{\bibfnamefont{M.}~\bibnamefont{Ibarra}}, \bibnamefont{and}
  \bibinfo{author}{\bibfnamefont{J.}~\bibnamefont{Santamar\'{\i}a}},
  \bibinfo{journal}{Nanotoday} \textbf{\bibinfo{volume}{2}},
  \bibinfo{pages}{22} (\bibinfo{year}{2007}).

\bibitem[{\citenamefont{Sun et~al.}(2008)\citenamefont{Sun, Lee, and
  Zhang}}]{Sun_magnetic_general}
\bibinfo{author}{\bibfnamefont{C.}~\bibnamefont{Sun}},
  \bibinfo{author}{\bibfnamefont{J.}~\bibnamefont{Lee}}, \bibnamefont{and}
  \bibinfo{author}{\bibfnamefont{M.}~\bibnamefont{Zhang}},
  \bibinfo{journal}{Adv. Drug Deliv. Rev.} \textbf{\bibinfo{volume}{60}},
  \bibinfo{pages}{1252} (\bibinfo{year}{2008}).

\bibitem[{\citenamefont{Mahmoudi et~al.}(2011)\citenamefont{Mahmoudi, Sant,
  Wang, Laurent, and Sen}}]{Mahmoudi_magnetic_general}
\bibinfo{author}{\bibfnamefont{M.}~\bibnamefont{Mahmoudi}},
  \bibinfo{author}{\bibfnamefont{S.}~\bibnamefont{Sant}},
  \bibinfo{author}{\bibfnamefont{B.}~\bibnamefont{Wang}},
  \bibinfo{author}{\bibfnamefont{S.}~\bibnamefont{Laurent}}, \bibnamefont{and}
  \bibinfo{author}{\bibfnamefont{T.}~\bibnamefont{Sen}}, \bibinfo{journal}{Adv.
  Drug Deliv. Rev.} \textbf{\bibinfo{volume}{63}}, \bibinfo{pages}{24}
  (\bibinfo{year}{2011}).

\bibitem[{\citenamefont{Chomoucka et~al.}(2010)\citenamefont{Chomoucka,
  Drbohlavova, Huska, Adam, Kizek, and Hubalek}}]{Chomoucka_magnetic_general}
\bibinfo{author}{\bibfnamefont{J.}~\bibnamefont{Chomoucka}},
  \bibinfo{author}{\bibfnamefont{J.}~\bibnamefont{Drbohlavova}},
  \bibinfo{author}{\bibfnamefont{D.}~\bibnamefont{Huska}},
  \bibinfo{author}{\bibfnamefont{V.}~\bibnamefont{Adam}},
  \bibinfo{author}{\bibfnamefont{R.}~\bibnamefont{Kizek}}, \bibnamefont{and}
  \bibinfo{author}{\bibfnamefont{J.}~\bibnamefont{Hubalek}},
  \bibinfo{journal}{Pharmacol. Res.} \textbf{\bibinfo{volume}{62}},
  \bibinfo{pages}{144} (\bibinfo{year}{2010}).

\bibitem[{\citenamefont{Hu et~al.}(2014)\citenamefont{Hu, Zhang, and
  Chen}}]{Hu_magnetic_general}
\bibinfo{author}{\bibfnamefont{L.}~\bibnamefont{Hu}},
  \bibinfo{author}{\bibfnamefont{R.}~\bibnamefont{Zhang}}, \bibnamefont{and}
  \bibinfo{author}{\bibfnamefont{Q.}~\bibnamefont{Chen}},
  \bibinfo{journal}{Nanoscale} \textbf{\bibinfo{volume}{6}},
  \bibinfo{pages}{14064} (\bibinfo{year}{2014}).

\bibitem[{\citenamefont{Lasic}(1993)}]{Lasic_1}
\bibinfo{author}{\bibfnamefont{D.}~\bibnamefont{Lasic}},
  \emph{\bibinfo{title}{Liposomes: From Physics to Applications}}
  (\bibinfo{publisher}{Elsevier, Amsterdam}, \bibinfo{year}{1993}).

\bibitem[{\citenamefont{Lasic}(1995)}]{Lasic_2}
\bibinfo{author}{\bibfnamefont{D.}~\bibnamefont{Lasic}},
  \emph{\bibinfo{title}{{Handbook of Biological Physics, Vol. 1, Chap. 10 {\rm
  Edited by R. Lipowsky and E. Sackmann}}}} (\bibinfo{publisher}{Elsevier,
  Amsterdam}, \bibinfo{year}{1995}).

\bibitem[{Las(1998)}]{Lasic_3}
\emph{\bibinfo{title}{{Medical Applications of Liposomes {\rm Edited by D.D.
  Lasic and D. Papahadjopoulos}}}} (\bibinfo{publisher}{Elsevier, Amsterdam},
  \bibinfo{year}{1998}).

\bibitem[{\citenamefont{De~Cuyper and Joniau}(1988)}]{Cuyper_1_sintesis}
\bibinfo{author}{\bibfnamefont{M.}~\bibnamefont{De~Cuyper}} \bibnamefont{and}
  \bibinfo{author}{\bibfnamefont{M.}~\bibnamefont{Joniau}},
  \bibinfo{journal}{Eur. Biophys. J.} \textbf{\bibinfo{volume}{15}},
  \bibinfo{pages}{311} (\bibinfo{year}{1988}).

\bibitem[{\citenamefont{Martina et~al.}(2005)\citenamefont{Martina, Fortin,
  M{\'e}nager, Cl{\'e}ment, Barratt, Grabielle-Madelmont, Gazeau, Cabuil, and
  Lesieur}}]{Martina_sintesis}
\bibinfo{author}{\bibfnamefont{M.}~\bibnamefont{Martina}},
  \bibinfo{author}{\bibfnamefont{J.}~\bibnamefont{Fortin}},
  \bibinfo{author}{\bibfnamefont{C.}~\bibnamefont{M{\'e}nager}},
  \bibinfo{author}{\bibfnamefont{O.}~\bibnamefont{Cl{\'e}ment}},
  \bibinfo{author}{\bibfnamefont{G.}~\bibnamefont{Barratt}},
  \bibinfo{author}{\bibfnamefont{C.}~\bibnamefont{Grabielle-Madelmont}},
  \bibinfo{author}{\bibfnamefont{F.}~\bibnamefont{Gazeau}},
  \bibinfo{author}{\bibfnamefont{V.}~\bibnamefont{Cabuil}}, \bibnamefont{and}
  \bibinfo{author}{\bibfnamefont{S.}~\bibnamefont{Lesieur}},
  \bibinfo{journal}{J. Am. Chem. Soc.} \textbf{\bibinfo{volume}{127}},
  \bibinfo{pages}{10676} (\bibinfo{year}{2005}).

\bibitem[{\citenamefont{Sabat{\'e} et~al.}(2008)\citenamefont{Sabat{\'e},
  Barnadas-Rodr\'{\i}guez, Callejas-Fern{\'a}ndez, Hidalgo-{\'A}lvarez, and
  Estelrich}}]{Sabate_sintesis}
\bibinfo{author}{\bibfnamefont{R.}~\bibnamefont{Sabat{\'e}}},
  \bibinfo{author}{\bibfnamefont{R.}~\bibnamefont{Barnadas-Rodr\'{\i}guez}},
  \bibinfo{author}{\bibfnamefont{J.}~\bibnamefont{Callejas-Fern{\'a}ndez}},
  \bibinfo{author}{\bibfnamefont{R.}~\bibnamefont{Hidalgo-{\'A}lvarez}},
  \bibnamefont{and}
  \bibinfo{author}{\bibfnamefont{J.}~\bibnamefont{Estelrich}},
  \bibinfo{journal}{Int. J. Pharm.} \textbf{\bibinfo{volume}{347}},
  \bibinfo{pages}{156} (\bibinfo{year}{2008}).

\bibitem[{\citenamefont{Plassat et~al.}(2007)\citenamefont{Plassat, Martina,
  Barratt, M{\'e}nager, and Lesieur}}]{Plassat_sintesis}
\bibinfo{author}{\bibfnamefont{V.}~\bibnamefont{Plassat}},
  \bibinfo{author}{\bibfnamefont{M.}~\bibnamefont{Martina}},
  \bibinfo{author}{\bibfnamefont{G.}~\bibnamefont{Barratt}},
  \bibinfo{author}{\bibfnamefont{C.}~\bibnamefont{M{\'e}nager}},
  \bibnamefont{and} \bibinfo{author}{\bibfnamefont{S.}~\bibnamefont{Lesieur}},
  \bibinfo{journal}{Int. J. Pharm.} \textbf{\bibinfo{volume}{344}},
  \bibinfo{pages}{118} (\bibinfo{year}{2007}).

\bibitem[{\citenamefont{Pereira~da Silva~Gomes
  et~al.}(2009)\citenamefont{Pereira~da Silva~Gomes, Rank, Kronenberger, Fritz,
  Winterhalter, and Ramaye}}]{Gomes_sintesis}
\bibinfo{author}{\bibfnamefont{J.}~\bibnamefont{Pereira~da Silva~Gomes}},
  \bibinfo{author}{\bibfnamefont{A.}~\bibnamefont{Rank}},
  \bibinfo{author}{\bibfnamefont{A.}~\bibnamefont{Kronenberger}},
  \bibinfo{author}{\bibfnamefont{J.}~\bibnamefont{Fritz}},
  \bibinfo{author}{\bibfnamefont{M.}~\bibnamefont{Winterhalter}},
  \bibnamefont{and} \bibinfo{author}{\bibfnamefont{Y.}~\bibnamefont{Ramaye}},
  \bibinfo{journal}{Langmuir} \textbf{\bibinfo{volume}{25}},
  \bibinfo{pages}{6793} (\bibinfo{year}{2009}).

\bibitem[{\citenamefont{Chen et~al.}(2010)\citenamefont{Chen, Bose, and
  Bothun}}]{Chen_sintesis}
\bibinfo{author}{\bibfnamefont{Y.}~\bibnamefont{Chen}},
  \bibinfo{author}{\bibfnamefont{A.}~\bibnamefont{Bose}}, \bibnamefont{and}
  \bibinfo{author}{\bibfnamefont{G.}~\bibnamefont{Bothun}},
  \bibinfo{journal}{ACS Nano} \textbf{\bibinfo{volume}{4}},
  \bibinfo{pages}{3215} (\bibinfo{year}{2010}).

\bibitem[{\citenamefont{Nappini et~al.}(2011)\citenamefont{Nappini, Bonini,
  Baldelli~Bombelli, Pineider, Sangregorio, Baglioni, and
  Nord{\`e}n}}]{Nappini_sintesis}
\bibinfo{author}{\bibfnamefont{S.}~\bibnamefont{Nappini}},
  \bibinfo{author}{\bibfnamefont{M.}~\bibnamefont{Bonini}},
  \bibinfo{author}{\bibfnamefont{F.}~\bibnamefont{Baldelli~Bombelli}},
  \bibinfo{author}{\bibfnamefont{F.}~\bibnamefont{Pineider}},
  \bibinfo{author}{\bibfnamefont{C.}~\bibnamefont{Sangregorio}},
  \bibinfo{author}{\bibfnamefont{P.}~\bibnamefont{Baglioni}}, \bibnamefont{and}
  \bibinfo{author}{\bibfnamefont{B.}~\bibnamefont{Nord{\`e}n}},
  \bibinfo{journal}{Soft Matter} \textbf{\bibinfo{volume}{7}},
  \bibinfo{pages}{1025} (\bibinfo{year}{2011}).

\bibitem[{\citenamefont{Garc{\'i}a-Jimeno
  et~al.}(2011)\citenamefont{Garc{\'i}a-Jimeno, Escribano, Queralt, and
  Estelrich}}]{Sonia_sintesis}
\bibinfo{author}{\bibfnamefont{S.}~\bibnamefont{Garc{\'i}a-Jimeno}},
  \bibinfo{author}{\bibfnamefont{E.}~\bibnamefont{Escribano}},
  \bibinfo{author}{\bibfnamefont{J.}~\bibnamefont{Queralt}}, \bibnamefont{and}
  \bibinfo{author}{\bibfnamefont{J.}~\bibnamefont{Estelrich}},
  \bibinfo{journal}{Int. J. Pharm.} \textbf{\bibinfo{volume}{405}},
  \bibinfo{pages}{181} (\bibinfo{year}{2011}).

\bibitem[{\citenamefont{Barenholz}(2001)}]{Barenholz}
\bibinfo{author}{\bibfnamefont{Y.}~\bibnamefont{Barenholz}},
  \bibinfo{journal}{Curr. Op. Interface Sci.} \textbf{\bibinfo{volume}{6}},
  \bibinfo{pages}{66} (\bibinfo{year}{2001}).

\bibitem[{\citenamefont{Hervault and Thanh}(2014)}]{Hervault_chemotherapy}
\bibinfo{author}{\bibfnamefont{A.}~\bibnamefont{Hervault}} \bibnamefont{and}
  \bibinfo{author}{\bibfnamefont{N.}~\bibnamefont{Thanh}},
  \bibinfo{journal}{Nanoscale} \textbf{\bibinfo{volume}{6}},
  \bibinfo{pages}{11553} (\bibinfo{year}{2014}).

\bibitem[{\citenamefont{Safarikova and
  Safarik}(2001)}]{Safarikova_hyperthermia}
\bibinfo{author}{\bibfnamefont{M.}~\bibnamefont{Safarikova}} \bibnamefont{and}
  \bibinfo{author}{\bibfnamefont{I.}~\bibnamefont{Safarik}},
  \bibinfo{journal}{Magn. Electr.} \textbf{\bibinfo{volume}{10}},
  \bibinfo{pages}{223} (\bibinfo{year}{2001}).

\bibitem[{\citenamefont{Hamaguchi et~al.}(2003)\citenamefont{Hamaguchi, Tohnai,
  Ito, Mitsudo, Shigetomi, Ito, Honda, Kobayashi, and
  Ueda}}]{Hamaguchi_hyperthermia}
\bibinfo{author}{\bibfnamefont{S.}~\bibnamefont{Hamaguchi}},
  \bibinfo{author}{\bibfnamefont{I.}~\bibnamefont{Tohnai}},
  \bibinfo{author}{\bibfnamefont{A.}~\bibnamefont{Ito}},
  \bibinfo{author}{\bibfnamefont{K.}~\bibnamefont{Mitsudo}},
  \bibinfo{author}{\bibfnamefont{T.}~\bibnamefont{Shigetomi}},
  \bibinfo{author}{\bibfnamefont{M.}~\bibnamefont{Ito}},
  \bibinfo{author}{\bibfnamefont{H.}~\bibnamefont{Honda}},
  \bibinfo{author}{\bibfnamefont{T.}~\bibnamefont{Kobayashi}},
  \bibnamefont{and} \bibinfo{author}{\bibfnamefont{M.}~\bibnamefont{Ueda}},
  \bibinfo{journal}{Cancer Sci.} \textbf{\bibinfo{volume}{94}},
  \bibinfo{pages}{834} (\bibinfo{year}{2003}).

\bibitem[{\citenamefont{Ito et~al.}(2004)\citenamefont{Ito, Kuga, Honda,
  Kikkawa, Horiuchi, Watanabe, and Kobayashi}}]{Ito_hyperthermia}
\bibinfo{author}{\bibfnamefont{A.}~\bibnamefont{Ito}},
  \bibinfo{author}{\bibfnamefont{Y.}~\bibnamefont{Kuga}},
  \bibinfo{author}{\bibfnamefont{H.}~\bibnamefont{Honda}},
  \bibinfo{author}{\bibfnamefont{H.}~\bibnamefont{Kikkawa}},
  \bibinfo{author}{\bibfnamefont{A.}~\bibnamefont{Horiuchi}},
  \bibinfo{author}{\bibfnamefont{Y.}~\bibnamefont{Watanabe}}, \bibnamefont{and}
  \bibinfo{author}{\bibfnamefont{T.}~\bibnamefont{Kobayashi}},
  \bibinfo{journal}{Cancer Lett.} \textbf{\bibinfo{volume}{212}},
  \bibinfo{pages}{167} (\bibinfo{year}{2004}).

\bibitem[{\citenamefont{Gonzales and Krishnan}(2005)}]{Gonzales_hyperthermia}
\bibinfo{author}{\bibfnamefont{M.}~\bibnamefont{Gonzales}} \bibnamefont{and}
  \bibinfo{author}{\bibfnamefont{K.}~\bibnamefont{Krishnan}},
  \bibinfo{journal}{J. Magn. Magn. Mater.} \textbf{\bibinfo{volume}{293}},
  \bibinfo{pages}{265} (\bibinfo{year}{2005}).

\bibitem[{\citenamefont{B{\'e}alle et~al.}(2012)\citenamefont{B{\'e}alle,
  Di~Corato, Kolosnjaj-Tabi, Dupuis, Cl{\'e}ment, Gazeau, Wilhelm, and
  M{\'e}nager}}]{Bealle_hyperthermia}
\bibinfo{author}{\bibfnamefont{G.}~\bibnamefont{B{\'e}alle}},
  \bibinfo{author}{\bibfnamefont{R.}~\bibnamefont{Di~Corato}},
  \bibinfo{author}{\bibfnamefont{J.}~\bibnamefont{Kolosnjaj-Tabi}},
  \bibinfo{author}{\bibfnamefont{V.}~\bibnamefont{Dupuis}},
  \bibinfo{author}{\bibfnamefont{O.}~\bibnamefont{Cl{\'e}ment}},
  \bibinfo{author}{\bibfnamefont{F.}~\bibnamefont{Gazeau}},
  \bibinfo{author}{\bibfnamefont{C.}~\bibnamefont{Wilhelm}}, \bibnamefont{and}
  \bibinfo{author}{\bibfnamefont{C.}~\bibnamefont{M{\'e}nager}},
  \bibinfo{journal}{Langmuir} \textbf{\bibinfo{volume}{28}},
  \bibinfo{pages}{11834} (\bibinfo{year}{2012}).

\bibitem[{\citenamefont{Bulte and De~Cuyper}(2003)}]{Bulte_resonance_imaging}
\bibinfo{author}{\bibfnamefont{J.}~\bibnamefont{Bulte}} \bibnamefont{and}
  \bibinfo{author}{\bibfnamefont{M.}~\bibnamefont{De~Cuyper}},
  \bibinfo{journal}{Methods Enzymol.} \textbf{\bibinfo{volume}{373}},
  \bibinfo{pages}{175} (\bibinfo{year}{2003}).

\bibitem[{\citenamefont{Du et~al.}(2015)\citenamefont{Du, Han, Li, Zhao, Su,
  Cao, and Zhang}}]{Du_resonance_imaging}
\bibinfo{author}{\bibfnamefont{B.}~\bibnamefont{Du}},
  \bibinfo{author}{\bibfnamefont{S.}~\bibnamefont{Han}},
  \bibinfo{author}{\bibfnamefont{H.}~\bibnamefont{Li}},
  \bibinfo{author}{\bibfnamefont{F.}~\bibnamefont{Zhao}},
  \bibinfo{author}{\bibfnamefont{X.}~\bibnamefont{Su}},
  \bibinfo{author}{\bibfnamefont{X.}~\bibnamefont{Cao}}, \bibnamefont{and}
  \bibinfo{author}{\bibfnamefont{Z.}~\bibnamefont{Zhang}},
  \bibinfo{journal}{Nanoscale} \textbf{\bibinfo{volume}{7}},
  \bibinfo{pages}{5411} (\bibinfo{year}{2015}).

\bibitem[{\citenamefont{Dandamudi and
  Campbell}(2006)}]{Dandamundi_magnetic_targeting}
\bibinfo{author}{\bibfnamefont{S.}~\bibnamefont{Dandamudi}} \bibnamefont{and}
  \bibinfo{author}{\bibfnamefont{R.}~\bibnamefont{Campbell}},
  \bibinfo{journal}{Biochim. Biophys. Acta} \textbf{\bibinfo{volume}{1768}},
  \bibinfo{pages}{427} (\bibinfo{year}{2006}).

\bibitem[{\citenamefont{Martina et~al.}(2008)\citenamefont{Martina, Wilhelm,
  and Lesieur}}]{Martina_magnetic_targeting}
\bibinfo{author}{\bibfnamefont{M.}~\bibnamefont{Martina}},
  \bibinfo{author}{\bibfnamefont{C.}~\bibnamefont{Wilhelm}}, \bibnamefont{and}
  \bibinfo{author}{\bibfnamefont{S.}~\bibnamefont{Lesieur}},
  \bibinfo{journal}{Biomaterials} \textbf{\bibinfo{volume}{29}},
  \bibinfo{pages}{4137} (\bibinfo{year}{2008}).

\bibitem[{\citenamefont{Soenen et~al.}(2011)\citenamefont{Soenen, Brisson,
  Jonckheere, Nuytten, Tan, Himmelreich, and
  De~Cuyper}}]{Soenen_magnetic_targeting}
\bibinfo{author}{\bibfnamefont{S.}~\bibnamefont{Soenen}},
  \bibinfo{author}{\bibfnamefont{A.}~\bibnamefont{Brisson}},
  \bibinfo{author}{\bibfnamefont{E.}~\bibnamefont{Jonckheere}},
  \bibinfo{author}{\bibfnamefont{N.}~\bibnamefont{Nuytten}},
  \bibinfo{author}{\bibfnamefont{S.}~\bibnamefont{Tan}},
  \bibinfo{author}{\bibfnamefont{U.}~\bibnamefont{Himmelreich}},
  \bibnamefont{and}
  \bibinfo{author}{\bibfnamefont{M.}~\bibnamefont{De~Cuyper}},
  \bibinfo{journal}{Biomaterials} \textbf{\bibinfo{volume}{32}},
  \bibinfo{pages}{1748} (\bibinfo{year}{2011}).

\bibitem[{\citenamefont{Benyettou et~al.}(2011)\citenamefont{Benyettou, Chebbi,
  Motte, and Seksek}}]{Benyettou_delivery}
\bibinfo{author}{\bibfnamefont{F.}~\bibnamefont{Benyettou}},
  \bibinfo{author}{\bibfnamefont{I.}~\bibnamefont{Chebbi}},
  \bibinfo{author}{\bibfnamefont{L.}~\bibnamefont{Motte}}, \bibnamefont{and}
  \bibinfo{author}{\bibfnamefont{O.}~\bibnamefont{Seksek}},
  \bibinfo{journal}{J. Mater. Chem.} \textbf{\bibinfo{volume}{21}},
  \bibinfo{pages}{4813} (\bibinfo{year}{2011}).

\bibitem[{\citenamefont{Saville et~al.}(2013)\citenamefont{Saville, Woodward,
  House, Tokarev, Hammers, Qi, Shaw, Saunders, Varsani, St~Pierre
  et~al.}}]{Saville_2013}
\bibinfo{author}{\bibfnamefont{S.}~\bibnamefont{Saville}},
  \bibinfo{author}{\bibfnamefont{R.}~\bibnamefont{Woodward}},
  \bibinfo{author}{\bibfnamefont{M.}~\bibnamefont{House}},
  \bibinfo{author}{\bibfnamefont{A.}~\bibnamefont{Tokarev}},
  \bibinfo{author}{\bibfnamefont{J.}~\bibnamefont{Hammers}},
  \bibinfo{author}{\bibfnamefont{B.}~\bibnamefont{Qi}},
  \bibinfo{author}{\bibfnamefont{J.}~\bibnamefont{Shaw}},
  \bibinfo{author}{\bibfnamefont{M.}~\bibnamefont{Saunders}},
  \bibinfo{author}{\bibfnamefont{R.}~\bibnamefont{Varsani}},
  \bibinfo{author}{\bibfnamefont{T.}~\bibnamefont{St~Pierre}},
  \bibnamefont{et~al.}, \bibinfo{journal}{Nanoscale}
  \textbf{\bibinfo{volume}{5}}, \bibinfo{pages}{2152} (\bibinfo{year}{2013}).

\bibitem[{\citenamefont{Saville et~al.}(2014)\citenamefont{Saville, Qi, Baker,
  Stone, Camley, Livesey, Ye, Crawford, and Mefford}}]{Saville_2014}
\bibinfo{author}{\bibfnamefont{S.}~\bibnamefont{Saville}},
  \bibinfo{author}{\bibfnamefont{B.}~\bibnamefont{Qi}},
  \bibinfo{author}{\bibfnamefont{J.}~\bibnamefont{Baker}},
  \bibinfo{author}{\bibfnamefont{R.}~\bibnamefont{Stone}},
  \bibinfo{author}{\bibfnamefont{R.}~\bibnamefont{Camley}},
  \bibinfo{author}{\bibfnamefont{K.}~\bibnamefont{Livesey}},
  \bibinfo{author}{\bibfnamefont{L.}~\bibnamefont{Ye}},
  \bibinfo{author}{\bibfnamefont{T.}~\bibnamefont{Crawford}}, \bibnamefont{and}
  \bibinfo{author}{\bibfnamefont{O.}~\bibnamefont{Mefford}},
  \bibinfo{journal}{J. Colloid Interface Sci.} \textbf{\bibinfo{volume}{424}},
  \bibinfo{pages}{141} (\bibinfo{year}{2014}).

\bibitem[{\citenamefont{Myrovali et~al.}(2016)\citenamefont{Myrovali, Maniotis,
  Makridis, Terzopoulou, Ntomprougkidis, Simeonidis, Sakellari, Kalogirou,
  Samaras, Salikhov et~al.}}]{Myrovali}
\bibinfo{author}{\bibfnamefont{E.}~\bibnamefont{Myrovali}},
  \bibinfo{author}{\bibfnamefont{N.}~\bibnamefont{Maniotis}},
  \bibinfo{author}{\bibfnamefont{A.}~\bibnamefont{Makridis}},
  \bibinfo{author}{\bibfnamefont{A.}~\bibnamefont{Terzopoulou}},
  \bibinfo{author}{\bibfnamefont{V.}~\bibnamefont{Ntomprougkidis}},
  \bibinfo{author}{\bibfnamefont{K.}~\bibnamefont{Simeonidis}},
  \bibinfo{author}{\bibfnamefont{D.}~\bibnamefont{Sakellari}},
  \bibinfo{author}{\bibfnamefont{O.}~\bibnamefont{Kalogirou}},
  \bibinfo{author}{\bibfnamefont{T.}~\bibnamefont{Samaras}},
  \bibinfo{author}{\bibfnamefont{R.}~\bibnamefont{Salikhov}},
  \bibnamefont{et~al.}, \bibinfo{journal}{Sci. Rep.}
  \textbf{\bibinfo{volume}{6}}, \bibinfo{pages}{37934} (\bibinfo{year}{2016}).

\bibitem[{\citenamefont{Jun et~al.}(2011)\citenamefont{Jun, Kim, Baek, Kang,
  Kim, Hyeon, Jeong, and Lee}}]{Jun}
\bibinfo{author}{\bibfnamefont{B.}~\bibnamefont{Jun}},
  \bibinfo{author}{\bibfnamefont{G.}~\bibnamefont{Kim}},
  \bibinfo{author}{\bibfnamefont{J.}~\bibnamefont{Baek}},
  \bibinfo{author}{\bibfnamefont{H.}~\bibnamefont{Kang}},
  \bibinfo{author}{\bibfnamefont{T.}~\bibnamefont{Kim}},
  \bibinfo{author}{\bibfnamefont{T.}~\bibnamefont{Hyeon}},
  \bibinfo{author}{\bibfnamefont{D.}~\bibnamefont{Jeong}}, \bibnamefont{and}
  \bibinfo{author}{\bibfnamefont{Y.}~\bibnamefont{Lee}},
  \bibinfo{journal}{Phys. Chem. Chem. Phys.} \textbf{\bibinfo{volume}{13}},
  \bibinfo{pages}{7298} (\bibinfo{year}{2011}).

\bibitem[{\citenamefont{Matsumura and Maeda}(1986)}]{Matsumura}
\bibinfo{author}{\bibfnamefont{Y.}~\bibnamefont{Matsumura}} \bibnamefont{and}
  \bibinfo{author}{\bibfnamefont{H.}~\bibnamefont{Maeda}},
  \bibinfo{journal}{Cancer Research} \textbf{\bibinfo{volume}{46}},
  \bibinfo{pages}{6387} (\bibinfo{year}{1986}).

\bibitem[{\citenamefont{Mody et~al.}(2014)\citenamefont{Mody, Cox, Shah, Singh,
  Bevins, and Parihar}}]{Application_aggregation_1}
\bibinfo{author}{\bibfnamefont{V.}~\bibnamefont{Mody}},
  \bibinfo{author}{\bibfnamefont{A.}~\bibnamefont{Cox}},
  \bibinfo{author}{\bibfnamefont{S.}~\bibnamefont{Shah}},
  \bibinfo{author}{\bibfnamefont{A.}~\bibnamefont{Singh}},
  \bibinfo{author}{\bibfnamefont{W.}~\bibnamefont{Bevins}}, \bibnamefont{and}
  \bibinfo{author}{\bibfnamefont{H.}~\bibnamefont{Parihar}},
  \bibinfo{journal}{Appl. Nanosci.} \textbf{\bibinfo{volume}{4}},
  \bibinfo{pages}{385} (\bibinfo{year}{2014}).

\bibitem[{\citenamefont{Licinio and Fr{\'e}zard}(2001)}]{Licinio}
\bibinfo{author}{\bibfnamefont{P.}~\bibnamefont{Licinio}} \bibnamefont{and}
  \bibinfo{author}{\bibfnamefont{F.}~\bibnamefont{Fr{\'e}zard}},
  \bibinfo{journal}{Brazilian J. of Physics} \textbf{\bibinfo{volume}{31}},
  \bibinfo{pages}{356} (\bibinfo{year}{2001}).

\bibitem[{\citenamefont{Mart\'{\i}nez-Pedrero
  et~al.}(2005)\citenamefont{Mart\'{\i}nez-Pedrero, Tirado-Miranda, Schmitt,
  and Callejas-Fern{\'a}ndez}}]{Fernando_CSIA_2005}
\bibinfo{author}{\bibfnamefont{F.}~\bibnamefont{Mart\'{\i}nez-Pedrero}},
  \bibinfo{author}{\bibfnamefont{M.}~\bibnamefont{Tirado-Miranda}},
  \bibinfo{author}{\bibfnamefont{A.}~\bibnamefont{Schmitt}}, \bibnamefont{and}
  \bibinfo{author}{\bibfnamefont{J.}~\bibnamefont{Callejas-Fern{\'a}ndez}},
  \bibinfo{journal}{Colloids Surf. A Physicochem Eng. Asp.}
  \textbf{\bibinfo{volume}{270}}, \bibinfo{pages}{317} (\bibinfo{year}{2005}).

\bibitem[{\citenamefont{Mart\'{\i}nez-Pedrero
  et~al.}(2006)\citenamefont{Mart\'{\i}nez-Pedrero, Tirado-Miranda, Schmitt,
  and Callejas-Fern{\'a}ndez}}]{Fernando_JCP_2006}
\bibinfo{author}{\bibfnamefont{F.}~\bibnamefont{Mart\'{\i}nez-Pedrero}},
  \bibinfo{author}{\bibfnamefont{M.}~\bibnamefont{Tirado-Miranda}},
  \bibinfo{author}{\bibfnamefont{A.}~\bibnamefont{Schmitt}}, \bibnamefont{and}
  \bibinfo{author}{\bibfnamefont{J.}~\bibnamefont{Callejas-Fern{\'a}ndez}},
  \bibinfo{journal}{J. Chem. Phys.} \textbf{\bibinfo{volume}{125}},
  \bibinfo{pages}{084706} (\bibinfo{year}{2006}).

\bibitem[{\citenamefont{Mart\'{\i}nez-Pedrero
  et~al.}(2007)\citenamefont{Mart\'{\i}nez-Pedrero, Tirado-Miranda, Schmitt,
  and Callejas-Fern{\'a}ndez}}]{Fernando_PRE_2007}
\bibinfo{author}{\bibfnamefont{F.}~\bibnamefont{Mart\'{\i}nez-Pedrero}},
  \bibinfo{author}{\bibfnamefont{M.}~\bibnamefont{Tirado-Miranda}},
  \bibinfo{author}{\bibfnamefont{A.}~\bibnamefont{Schmitt}}, \bibnamefont{and}
  \bibinfo{author}{\bibfnamefont{J.}~\bibnamefont{Callejas-Fern{\'a}ndez}},
  \bibinfo{journal}{Phys. Rev. E} \textbf{\bibinfo{volume}{76}},
  \bibinfo{pages}{011405} (\bibinfo{year}{2007}).

\bibitem[{\citenamefont{Mart\'{\i}nez-Pedrero
  et~al.}(2008{\natexlab{a}})\citenamefont{Mart\'{\i}nez-Pedrero,
  Tirado-Miranda, Schmitt, and Callejas-Fern{\'a}ndez}}]{Fernando_JCIS_2008}
\bibinfo{author}{\bibfnamefont{F.}~\bibnamefont{Mart\'{\i}nez-Pedrero}},
  \bibinfo{author}{\bibfnamefont{M.}~\bibnamefont{Tirado-Miranda}},
  \bibinfo{author}{\bibfnamefont{A.}~\bibnamefont{Schmitt}}, \bibnamefont{and}
  \bibinfo{author}{\bibfnamefont{J.}~\bibnamefont{Callejas-Fern{\'a}ndez}},
  \bibinfo{journal}{‎J. Colloid Interface Sci.}
  \textbf{\bibinfo{volume}{318}}, \bibinfo{pages}{23}
  (\bibinfo{year}{2008}{\natexlab{a}}).

\bibitem[{\citenamefont{Mart\'{\i}nez-Pedrero
  et~al.}(2009)\citenamefont{Mart\'{\i}nez-Pedrero, Tirado-Miranda, Schmitt,
  and Callejas-Fern{\'a}ndez}}]{Fernando_Langmuir_2009}
\bibinfo{author}{\bibfnamefont{F.}~\bibnamefont{Mart\'{\i}nez-Pedrero}},
  \bibinfo{author}{\bibfnamefont{M.}~\bibnamefont{Tirado-Miranda}},
  \bibinfo{author}{\bibfnamefont{A.}~\bibnamefont{Schmitt}}, \bibnamefont{and}
  \bibinfo{author}{\bibfnamefont{J.}~\bibnamefont{Callejas-Fern{\'a}ndez}},
  \bibinfo{journal}{Langmuir} \textbf{\bibinfo{volume}{25}},
  \bibinfo{pages}{6658} (\bibinfo{year}{2009}).

\bibitem[{\citenamefont{Dom\'{\i}nguez-Garc\'{\i}a and
  Rubio}(2010)}]{Dominguez_2_morphology}
\bibinfo{author}{\bibfnamefont{P.}~\bibnamefont{Dom\'{\i}nguez-Garc\'{\i}a}}
  \bibnamefont{and} \bibinfo{author}{\bibfnamefont{M.}~\bibnamefont{Rubio}},
  \bibinfo{journal}{Colloids Surf. A Physicochem Eng. Asp.}
  \textbf{\bibinfo{volume}{358}}, \bibinfo{pages}{21} (\bibinfo{year}{2010}).

\bibitem[{\citenamefont{Mart\'{\i}nez-Pedrero
  et~al.}(2008{\natexlab{b}})\citenamefont{Mart\'{\i}nez-Pedrero, El-Harrak,
  Fern{\'a}ndez-Toledano, Tirado-Miranda, Baudry, Schmitt, Bibette, and
  Callejas-Fern{\'a}ndez}}]{Fernando_PRE_2008}
\bibinfo{author}{\bibfnamefont{F.}~\bibnamefont{Mart\'{\i}nez-Pedrero}},
  \bibinfo{author}{\bibfnamefont{A.}~\bibnamefont{El-Harrak}},
  \bibinfo{author}{\bibfnamefont{J.}~\bibnamefont{Fern{\'a}ndez-Toledano}},
  \bibinfo{author}{\bibfnamefont{M.}~\bibnamefont{Tirado-Miranda}},
  \bibinfo{author}{\bibfnamefont{J.}~\bibnamefont{Baudry}},
  \bibinfo{author}{\bibfnamefont{A.}~\bibnamefont{Schmitt}},
  \bibinfo{author}{\bibfnamefont{J.}~\bibnamefont{Bibette}}, \bibnamefont{and}
  \bibinfo{author}{\bibfnamefont{J.}~\bibnamefont{Callejas-Fern{\'a}ndez}},
  \bibinfo{journal}{Phys. Rev. E} \textbf{\bibinfo{volume}{78}},
  \bibinfo{pages}{011403} (\bibinfo{year}{2008}{\natexlab{b}}).

\bibitem[{\citenamefont{Dom\'{\i}nguez-Garc\'{\i}a
  et~al.}(2009)\citenamefont{Dom\'{\i}nguez-Garc\'{\i}a, Melle, and
  Rubio}}]{Dominguez_1_morphology}
\bibinfo{author}{\bibfnamefont{P.}~\bibnamefont{Dom\'{\i}nguez-Garc\'{\i}a}},
  \bibinfo{author}{\bibfnamefont{S.}~\bibnamefont{Melle}}, \bibnamefont{and}
  \bibinfo{author}{\bibfnamefont{M.}~\bibnamefont{Rubio}}, \bibinfo{journal}{J.
  Colloid Interface Sci.} \textbf{\bibinfo{volume}{333}}, \bibinfo{pages}{221}
  (\bibinfo{year}{2009}).

\bibitem[{\citenamefont{Andreu et~al.}(2011)\citenamefont{Andreu, Camacho, and
  Faraudo}}]{Faraudo_softmatter}
\bibinfo{author}{\bibfnamefont{J.~S.} \bibnamefont{Andreu}},
  \bibinfo{author}{\bibfnamefont{J.}~\bibnamefont{Camacho}}, \bibnamefont{and}
  \bibinfo{author}{\bibfnamefont{J.}~\bibnamefont{Faraudo}},
  \bibinfo{journal}{Soft Matter} \textbf{\bibinfo{volume}{7}},
  \bibinfo{pages}{2336} (\bibinfo{year}{2011}).

\bibitem[{\citenamefont{Bertoni et~al.}(2011)\citenamefont{Bertoni, Torre,
  Falqui, Fragouli, Athanassiou, and Cingolani}}]{Bertoni_simulation}
\bibinfo{author}{\bibfnamefont{G.}~\bibnamefont{Bertoni}},
  \bibinfo{author}{\bibfnamefont{B.}~\bibnamefont{Torre}},
  \bibinfo{author}{\bibfnamefont{A.}~\bibnamefont{Falqui}},
  \bibinfo{author}{\bibfnamefont{D.}~\bibnamefont{Fragouli}},
  \bibinfo{author}{\bibfnamefont{A.}~\bibnamefont{Athanassiou}},
  \bibnamefont{and}
  \bibinfo{author}{\bibfnamefont{R.}~\bibnamefont{Cingolani}},
  \bibinfo{journal}{J. Phys. Chem. C} \textbf{\bibinfo{volume}{115}},
  \bibinfo{pages}{7249} (\bibinfo{year}{2011}).

\bibitem[{\citenamefont{Rovigatti et~al.}(2011)\citenamefont{Rovigatti, Russo,
  and Sciortino}}]{Lorenzo_rings}
\bibinfo{author}{\bibfnamefont{L.}~\bibnamefont{Rovigatti}},
  \bibinfo{author}{\bibfnamefont{J.}~\bibnamefont{Russo}}, \bibnamefont{and}
  \bibinfo{author}{\bibfnamefont{F.}~\bibnamefont{Sciortino}},
  \bibinfo{journal}{Phys. Rev. Lett.} \textbf{\bibinfo{volume}{107}},
  \bibinfo{pages}{237801} (\bibinfo{year}{2011}).

\bibitem[{\citenamefont{Tlusty and Safran}(2000)}]{Tlusty_Safran}
\bibinfo{author}{\bibfnamefont{T.}~\bibnamefont{Tlusty}} \bibnamefont{and}
  \bibinfo{author}{\bibfnamefont{S.~A.} \bibnamefont{Safran}},
  \bibinfo{journal}{Science} \textbf{\bibinfo{volume}{290}},
  \bibinfo{pages}{1328} (\bibinfo{year}{2000}).

\bibitem[{\citenamefont{Cerd{\`a} et~al.}(2008)\citenamefont{Cerd{\`a},
  Kantorovich, and Holm}}]{Cerda_1_simulation}
\bibinfo{author}{\bibfnamefont{J.}~\bibnamefont{Cerd{\`a}}},
  \bibinfo{author}{\bibfnamefont{S.}~\bibnamefont{Kantorovich}},
  \bibnamefont{and} \bibinfo{author}{\bibfnamefont{C.}~\bibnamefont{Holm}},
  \bibinfo{journal}{J. Phys.: Condens. Matter} \textbf{\bibinfo{volume}{20}},
  \bibinfo{pages}{204125} (\bibinfo{year}{2008}).

\bibitem[{\citenamefont{Szoka and
  Papahadjopoulos}(1978)}]{Szoka-Papahadjopoulos}
\bibinfo{author}{\bibfnamefont{F.}~\bibnamefont{Szoka}} \bibnamefont{and}
  \bibinfo{author}{\bibfnamefont{D.}~\bibnamefont{Papahadjopoulos}},
  \bibinfo{journal}{Proc. Natl. Acad. Sci. USA} \textbf{\bibinfo{volume}{75}},
  \bibinfo{pages}{4194} (\bibinfo{year}{1978}).

\bibitem[{\citenamefont{MacDonald et~al.}(1991)\citenamefont{MacDonald,
  MacDonald, Menco, Takeshita, Subbarao, and Hu}}]{MacDonald}
\bibinfo{author}{\bibfnamefont{R.~C.} \bibnamefont{MacDonald}},
  \bibinfo{author}{\bibfnamefont{R.~I.} \bibnamefont{MacDonald}},
  \bibinfo{author}{\bibfnamefont{B.~P.} \bibnamefont{Menco}},
  \bibinfo{author}{\bibfnamefont{K.}~\bibnamefont{Takeshita}},
  \bibinfo{author}{\bibfnamefont{N.~K.} \bibnamefont{Subbarao}},
  \bibnamefont{and} \bibinfo{author}{\bibfnamefont{L.~R.} \bibnamefont{Hu}},
  \bibinfo{journal}{Biochim. Biophys. Acta} \textbf{\bibinfo{volume}{1061}},
  \bibinfo{pages}{297} (\bibinfo{year}{1991}).

\bibitem[{\citenamefont{Steward-Marshall}(1980)}]{Steward-Marshall}
\bibinfo{author}{\bibfnamefont{J.~C.} \bibnamefont{Steward-Marshall}},
  \bibinfo{journal}{Anal. Biochem.} \textbf{\bibinfo{volume}{104}},
  \bibinfo{pages}{10} (\bibinfo{year}{1980}).

\bibitem[{\citenamefont{Pusey}(1989)}]{pusey_1989}
\bibinfo{author}{\bibfnamefont{P.}~\bibnamefont{Pusey}},
  \emph{\bibinfo{title}{Liquids, Freezing and Glass Transition, Lecture Notes
  for Les Houches, Session LI, Part 2, Course 10: Colloidal Suspensions.}}
  (\bibinfo{publisher}{North Holland, Amsterdam}, \bibinfo{year}{1989}).

\bibitem[{\citenamefont{Pusey}(2002)}]{pusey_2002}
\bibinfo{author}{\bibfnamefont{P.}~\bibnamefont{Pusey}},
  \emph{\bibinfo{title}{Introduction to scattering experiments (in Neutrons,
  X-rays and light: scattering methods applied to soft condensed matter)}}
  (\bibinfo{publisher}{North Holland, Elsevier, Amsterdam},
  \bibinfo{year}{2002}).

\bibitem[{\citenamefont{Rold\'an-Vargas
  et~al.}(2007)\citenamefont{Rold\'an-Vargas, Mart\'{\i}n-Molina,
  Quesada-P\'erez, Barnadas-Rodr\'{\i}guez, Estelrich, and
  Callejas-Fern\'andez}}]{calcio}
\bibinfo{author}{\bibfnamefont{S.}~\bibnamefont{Rold\'an-Vargas}},
  \bibinfo{author}{\bibfnamefont{A.}~\bibnamefont{Mart\'{\i}n-Molina}},
  \bibinfo{author}{\bibfnamefont{M.}~\bibnamefont{Quesada-P\'erez}},
  \bibinfo{author}{\bibfnamefont{R.}~\bibnamefont{Barnadas-Rodr\'{\i}guez}},
  \bibinfo{author}{\bibfnamefont{J.}~\bibnamefont{Estelrich}},
  \bibnamefont{and}
  \bibinfo{author}{\bibfnamefont{J.}~\bibnamefont{Callejas-Fern\'andez}},
  \bibinfo{journal}{Phys. Rev. E} \textbf{\bibinfo{volume}{75}},
  \bibinfo{pages}{021912} (\bibinfo{year}{2007}).

\bibitem[{\citenamefont{Rold\'an-Vargas
  et~al.}(2009{\natexlab{a}})\citenamefont{Rold\'an-Vargas,
  Barnadas-Rodr\'{\i}guez, Quesada-P\'erez, Estelrich, and
  Callejas-Fern\'andez}}]{surface}
\bibinfo{author}{\bibfnamefont{S.}~\bibnamefont{Rold\'an-Vargas}},
  \bibinfo{author}{\bibfnamefont{R.}~\bibnamefont{Barnadas-Rodr\'{\i}guez}},
  \bibinfo{author}{\bibfnamefont{M.}~\bibnamefont{Quesada-P\'erez}},
  \bibinfo{author}{\bibfnamefont{J.}~\bibnamefont{Estelrich}},
  \bibnamefont{and}
  \bibinfo{author}{\bibfnamefont{J.}~\bibnamefont{Callejas-Fern\'andez}},
  \bibinfo{journal}{Phys. Rev. E} \textbf{\bibinfo{volume}{79}},
  \bibinfo{pages}{011905} (\bibinfo{year}{2009}{\natexlab{a}}).

\bibitem[{\citenamefont{Sorensen}(2001)}]{sorensen}
\bibinfo{author}{\bibfnamefont{C.}~\bibnamefont{Sorensen}},
  \bibinfo{journal}{Aerosol Sci. Technol.} \textbf{\bibinfo{volume}{35}},
  \bibinfo{pages}{648} (\bibinfo{year}{2001}).

\bibitem[{\citenamefont{Berne and Pecora}(2000)}]{berne_pecora}
\bibinfo{author}{\bibfnamefont{B.}~\bibnamefont{Berne}} \bibnamefont{and}
  \bibinfo{author}{\bibfnamefont{R.}~\bibnamefont{Pecora}},
  \emph{\bibinfo{title}{Dynamic Light Scattering With Applications to
  Chemistry, Biology, and Physics}} (\bibinfo{publisher}{Dover Publications,
  Mineola, New York}, \bibinfo{year}{2000}).

\bibitem[{\citenamefont{Koppel}(1972)}]{koppel}
\bibinfo{author}{\bibfnamefont{D.}~\bibnamefont{Koppel}}, \bibinfo{journal}{J.
  Chem. Phys.} \textbf{\bibinfo{volume}{57}}, \bibinfo{pages}{4814}
  (\bibinfo{year}{1972}).

\bibitem[{\citenamefont{Dhont}(1996)}]{Dhont}
\bibinfo{author}{\bibfnamefont{J.}~\bibnamefont{Dhont}},
  \emph{\bibinfo{title}{An Introduction to Dynamics of Colloids}}
  (\bibinfo{publisher}{Amsterdam: Elsevier}, \bibinfo{year}{1996}).

\bibitem[{\citenamefont{Rold\'an-Vargas
  et~al.}(2009{\natexlab{b}})\citenamefont{Rold\'an-Vargas, Quesada-P\'erez,
  and Callejas-Fern\'andez}}]{stochastic}
\bibinfo{author}{\bibfnamefont{S.}~\bibnamefont{Rold\'an-Vargas}},
  \bibinfo{author}{\bibfnamefont{M.}~\bibnamefont{Quesada-P\'erez}},
  \bibnamefont{and}
  \bibinfo{author}{\bibfnamefont{J.}~\bibnamefont{Callejas-Fern\'andez}},
  \bibinfo{journal}{J. Chem. Phys.} \textbf{\bibinfo{volume}{131}},
  \bibinfo{pages}{034509} (\bibinfo{year}{2009}{\natexlab{b}}).

\bibitem[{\citenamefont{Derjaguin and Landau}(1941)}]{DLVO_1}
\bibinfo{author}{\bibfnamefont{B.~V.} \bibnamefont{Derjaguin}}
  \bibnamefont{and} \bibinfo{author}{\bibfnamefont{L.}~\bibnamefont{Landau}},
  \bibinfo{journal}{Acta Physicochim. URSS} \textbf{\bibinfo{volume}{14}},
  \bibinfo{pages}{633} (\bibinfo{year}{1941}).

\bibitem[{\citenamefont{Verwey and Overbeek}(1948)}]{DLVO_2}
\bibinfo{author}{\bibfnamefont{E.}~\bibnamefont{Verwey}} \bibnamefont{and}
  \bibinfo{author}{\bibfnamefont{J.}~\bibnamefont{Overbeek}},
  \emph{\bibinfo{title}{Theory of the Stability of Lyophobic Colloids}}
  (\bibinfo{publisher}{Elsevier, Amsterdam}, \bibinfo{year}{1948}).

\bibitem[{\citenamefont{T.J.}(1996)}]{hydration_PC}
\bibinfo{author}{\bibfnamefont{M.}~\bibnamefont{T.J.}}, \bibinfo{journal}{Chem
  Phys Lipids} \textbf{\bibinfo{volume}{81}}, \bibinfo{pages}{117}
  (\bibinfo{year}{1996}).

\bibitem[{\citenamefont{Israelachvili and
  Wennerstrom}(1996)}]{Israelachvili_Wennerstrom}
\bibinfo{author}{\bibfnamefont{J.}~\bibnamefont{Israelachvili}}
  \bibnamefont{and}
  \bibinfo{author}{\bibfnamefont{H.}~\bibnamefont{Wennerstrom}},
  \bibinfo{journal}{Nature} \textbf{\bibinfo{volume}{379}},
  \bibinfo{pages}{219} (\bibinfo{year}{1996}).

\bibitem[{\citenamefont{Petsev and Vekilov}(2000)}]{Petsev_Vekilov}
\bibinfo{author}{\bibfnamefont{D.}~\bibnamefont{Petsev}} \bibnamefont{and}
  \bibinfo{author}{\bibfnamefont{P.}~\bibnamefont{Vekilov}},
  \bibinfo{journal}{Chem. Phys. Lett.} \textbf{\bibinfo{volume}{84}},
  \bibinfo{pages}{1339} (\bibinfo{year}{2000}).

\bibitem[{\citenamefont{Marcella}(1976)}]{Marcella}
\bibinfo{author}{\bibfnamefont{S.}~\bibnamefont{Marcella}},
  \bibinfo{journal}{Phys. Rev. Lett.} \textbf{\bibinfo{volume}{42}},
  \bibinfo{pages}{129} (\bibinfo{year}{1976}).

\bibitem[{\citenamefont{Ohki and Arnold}(2000)}]{Ohki_Arnold}
\bibinfo{author}{\bibfnamefont{S.}~\bibnamefont{Ohki}} \bibnamefont{and}
  \bibinfo{author}{\bibfnamefont{K.}~\bibnamefont{Arnold}},
  \bibinfo{journal}{Colloids Surf. B} \textbf{\bibinfo{volume}{18}},
  \bibinfo{pages}{83} (\bibinfo{year}{2000}).

\bibitem[{\citenamefont{LeNeveu et~al.}(1976)\citenamefont{LeNeveu, Rand, and
  Parsegian}}]{LeNeveu}
\bibinfo{author}{\bibfnamefont{D.}~\bibnamefont{LeNeveu}},
  \bibinfo{author}{\bibfnamefont{P.}~\bibnamefont{Rand}}, \bibnamefont{and}
  \bibinfo{author}{\bibfnamefont{V.}~\bibnamefont{Parsegian}},
  \bibinfo{journal}{Nature} \textbf{\bibinfo{volume}{259}},
  \bibinfo{pages}{601} (\bibinfo{year}{1976}).

\bibitem[{\citenamefont{Chan et~al.}(1985)\citenamefont{Chan, Henderson,
  Barojas, and Homola}}]{magnetic_interaction_1}
\bibinfo{author}{\bibfnamefont{Y.}~\bibnamefont{Chan}},
  \bibinfo{author}{\bibfnamefont{D.}~\bibnamefont{Henderson}},
  \bibinfo{author}{\bibfnamefont{J.}~\bibnamefont{Barojas}}, \bibnamefont{and}
  \bibinfo{author}{\bibfnamefont{A.}~\bibnamefont{Homola}},
  \bibinfo{journal}{IBM J. Res. Develop.} \textbf{\bibinfo{volume}{29}},
  \bibinfo{pages}{11} (\bibinfo{year}{1985}).

\bibitem[{\citenamefont{Promislow and Gast}(1995)}]{magnetic_interaction_2}
\bibinfo{author}{\bibfnamefont{J.}~\bibnamefont{Promislow}} \bibnamefont{and}
  \bibinfo{author}{\bibfnamefont{A.}~\bibnamefont{Gast}}, \bibinfo{journal}{J.
  Chem. Phys.} \textbf{\bibinfo{volume}{102}}, \bibinfo{pages}{5942}
  (\bibinfo{year}{1995}).

\bibitem[{\citenamefont{Tsouris and Scott}(1995)}]{magnetic_interaction_3}
\bibinfo{author}{\bibfnamefont{C.}~\bibnamefont{Tsouris}} \bibnamefont{and}
  \bibinfo{author}{\bibfnamefont{T.}~\bibnamefont{Scott}}, \bibinfo{journal}{J.
  Colloid Interface Sci.} \textbf{\bibinfo{volume}{171}}, \bibinfo{pages}{319}
  (\bibinfo{year}{1995}).

\bibitem[{\citenamefont{Chikazumi}(1964)}]{chikazumi}
\bibinfo{author}{\bibfnamefont{S.}~\bibnamefont{Chikazumi}},
  \emph{\bibinfo{title}{Physics of Magnetism}} (\bibinfo{publisher}{Wiley, New
  York}, \bibinfo{year}{1964}).

\bibitem[{\citenamefont{Leyvraz}(2003)}]{leyvraz}
\bibinfo{author}{\bibfnamefont{F.}~\bibnamefont{Leyvraz}},
  \bibinfo{journal}{Phys. Rep.} \textbf{\bibinfo{volume}{383}},
  \bibinfo{pages}{95} (\bibinfo{year}{2003}).

\bibitem[{\citenamefont{Kolb}(1984)}]{kolb}
\bibinfo{author}{\bibfnamefont{M.}~\bibnamefont{Kolb}}, \bibinfo{journal}{Phys.
  Rev. Lett.} \textbf{\bibinfo{volume}{53}}, \bibinfo{pages}{1653}
  (\bibinfo{year}{1984}).

\bibitem[{\citenamefont{Weitz and Huang}(1984)}]{weitz_huang}
\bibinfo{author}{\bibfnamefont{D.}~\bibnamefont{Weitz}} \bibnamefont{and}
  \bibinfo{author}{\bibfnamefont{J.}~\bibnamefont{Huang}},
  \emph{\bibinfo{title}{Self-Similar Structures and the Kinetics of Aggregation
  of Gold Colloids}} (\bibinfo{publisher}{Elsevier Science, New York},
  \bibinfo{year}{1984}).

\bibitem[{\citenamefont{Lin et~al.}(1990)\citenamefont{Lin, Lindsay, Weitz,
  Klein, Ball, and Meakin}}]{lin}
\bibinfo{author}{\bibfnamefont{M.~Y.} \bibnamefont{Lin}},
  \bibinfo{author}{\bibfnamefont{H.~M.} \bibnamefont{Lindsay}},
  \bibinfo{author}{\bibfnamefont{D.~A.} \bibnamefont{Weitz}},
  \bibinfo{author}{\bibfnamefont{R.}~\bibnamefont{Klein}},
  \bibinfo{author}{\bibfnamefont{R.~C.} \bibnamefont{Ball}}, \bibnamefont{and}
  \bibinfo{author}{\bibfnamefont{P.}~\bibnamefont{Meakin}},
  \bibinfo{journal}{J. Phys. Condens. Matter} \textbf{\bibinfo{volume}{2}},
  \bibinfo{pages}{3093} (\bibinfo{year}{1990}).

\bibitem[{\citenamefont{van Dongen and
  Ernst}(1984)}]{Dongen_Ernst_Fragmentation}
\bibinfo{author}{\bibfnamefont{P.}~\bibnamefont{van Dongen}} \bibnamefont{and}
  \bibinfo{author}{\bibfnamefont{M.}~\bibnamefont{Ernst}}, \bibinfo{journal}{J.
  Stat. Phys.} \textbf{\bibinfo{volume}{37}}, \bibinfo{pages}{301}
  (\bibinfo{year}{1984}).

\bibitem[{\citenamefont{Family et~al.}(1986)\citenamefont{Family, Meakin, and
  Deutch}}]{Fragmentation_Family}
\bibinfo{author}{\bibfnamefont{F.}~\bibnamefont{Family}},
  \bibinfo{author}{\bibfnamefont{P.}~\bibnamefont{Meakin}}, \bibnamefont{and}
  \bibinfo{author}{\bibfnamefont{J.}~\bibnamefont{Deutch}},
  \bibinfo{journal}{Phys. Rev. Lett.} \textbf{\bibinfo{volume}{57}},
  \bibinfo{pages}{727} (\bibinfo{year}{1986}).

\bibitem[{\citenamefont{Rovigatti et~al.}(2012)\citenamefont{Rovigatti, Russo,
  and Sciortino}}]{Lorenzo_df}
\bibinfo{author}{\bibfnamefont{L.}~\bibnamefont{Rovigatti}},
  \bibinfo{author}{\bibfnamefont{J.}~\bibnamefont{Russo}}, \bibnamefont{and}
  \bibinfo{author}{\bibfnamefont{F.}~\bibnamefont{Sciortino}},
  \bibinfo{journal}{Soft Matter} \textbf{\bibinfo{volume}{8}},
  \bibinfo{pages}{6310} (\bibinfo{year}{2012}).

\bibitem[{\citenamefont{Camp and Patey}(2000)}]{Camp_Patey}
\bibinfo{author}{\bibfnamefont{P.~J.} \bibnamefont{Camp}} \bibnamefont{and}
  \bibinfo{author}{\bibfnamefont{G.~N.} \bibnamefont{Patey}},
  \bibinfo{journal}{Phys. Rev. E} \textbf{\bibinfo{volume}{62}},
  \bibinfo{pages}{5403} (\bibinfo{year}{2000}).

\bibitem[{\citenamefont{Brodie}(1981)}]{brodie}
\bibinfo{author}{\bibfnamefont{M.}~\bibnamefont{Brodie}},
  \emph{\bibinfo{title}{Experimental study of aggregation kinetics: dynamic
  scaling of measured Cluster-Size Distributions}}
  (\bibinfo{publisher}{Massachusetts Institute of Technology, PhD. thesis},
  \bibinfo{year}{1981}).

\bibitem[{\citenamefont{van Dongen and Ernst}(1985)}]{Dongen_Ernst_lambda_1}
\bibinfo{author}{\bibfnamefont{P.}~\bibnamefont{van Dongen}} \bibnamefont{and}
  \bibinfo{author}{\bibfnamefont{M.}~\bibnamefont{Ernst}},
  \bibinfo{journal}{Phys. Rev. Lett.} \textbf{\bibinfo{volume}{54}},
  \bibinfo{pages}{1396} (\bibinfo{year}{1985}).

\bibitem[{\citenamefont{van Dongen and Ernst}(1988)}]{Dongen_Ernst_lambda_2}
\bibinfo{author}{\bibfnamefont{P.}~\bibnamefont{van Dongen}} \bibnamefont{and}
  \bibinfo{author}{\bibfnamefont{M.}~\bibnamefont{Ernst}}, \bibinfo{journal}{J.
  Stat. Phys.} \textbf{\bibinfo{volume}{50}}, \bibinfo{pages}{295}
  (\bibinfo{year}{1988}).

\bibitem[{\citenamefont{Kantorovich and Ivanov}(2002)}]{Kantorovich_1}
\bibinfo{author}{\bibfnamefont{S.}~\bibnamefont{Kantorovich}} \bibnamefont{and}
  \bibinfo{author}{\bibfnamefont{A.~O.} \bibnamefont{Ivanov}},
  \bibinfo{journal}{‎J. Magn. Magn. Mater.} \textbf{\bibinfo{volume}{252}},
  \bibinfo{pages}{244} (\bibinfo{year}{2002}).

\bibitem[{\citenamefont{Ivanov and Kantorovich}(2004)}]{Kantorovich_2}
\bibinfo{author}{\bibfnamefont{A.~O.} \bibnamefont{Ivanov}} \bibnamefont{and}
  \bibinfo{author}{\bibfnamefont{S.}~\bibnamefont{Kantorovich}},
  \bibinfo{journal}{‎Phys. Rev. E} \textbf{\bibinfo{volume}{70}},
  \bibinfo{pages}{021401} (\bibinfo{year}{2004}).

\bibitem[{\citenamefont{Mart\'{\i}nez-Pedrero
  et~al.}(2016)\citenamefont{Mart\'{\i}nez-Pedrero, Cebers, and
  Tierno}}]{Fernando_rings}
\bibinfo{author}{\bibfnamefont{F.}~\bibnamefont{Mart\'{\i}nez-Pedrero}},
  \bibinfo{author}{\bibfnamefont{A.}~\bibnamefont{Cebers}}, \bibnamefont{and}
  \bibinfo{author}{\bibfnamefont{P.}~\bibnamefont{Tierno}},
  \bibinfo{journal}{Phys. Rev. Applied} \textbf{\bibinfo{volume}{6}},
  \bibinfo{pages}{034002} (\bibinfo{year}{2016}).

\bibitem[{\citenamefont{Chin et~al.}(2001)\citenamefont{Chin, Yiacoumi, and
  Tsouris}}]{primary_secondary_minima_1}
\bibinfo{author}{\bibfnamefont{C.}~\bibnamefont{Chin}},
  \bibinfo{author}{\bibfnamefont{S.}~\bibnamefont{Yiacoumi}}, \bibnamefont{and}
  \bibinfo{author}{\bibfnamefont{C.}~\bibnamefont{Tsouris}},
  \bibinfo{journal}{Langmuir} \textbf{\bibinfo{volume}{17}},
  \bibinfo{pages}{6065} (\bibinfo{year}{2001}).

\bibitem[{\citenamefont{Yiacoumi et~al.}(1996)\citenamefont{Yiacoumi, Rountree,
  and Tsouris}}]{primary_secondary_minima_2}
\bibinfo{author}{\bibfnamefont{S.}~\bibnamefont{Yiacoumi}},
  \bibinfo{author}{\bibfnamefont{D.}~\bibnamefont{Rountree}}, \bibnamefont{and}
  \bibinfo{author}{\bibfnamefont{C.}~\bibnamefont{Tsouris}},
  \bibinfo{journal}{J. Colloid Interface Sci.} \textbf{\bibinfo{volume}{184}},
  \bibinfo{pages}{477} (\bibinfo{year}{1996}).

\bibitem[{\citenamefont{Chin et~al.}(1998)\citenamefont{Chin, Yiacoumi, and
  Tsouris}}]{primary_secondary_minima_3}
\bibinfo{author}{\bibfnamefont{C.}~\bibnamefont{Chin}},
  \bibinfo{author}{\bibfnamefont{S.}~\bibnamefont{Yiacoumi}}, \bibnamefont{and}
  \bibinfo{author}{\bibfnamefont{C.}~\bibnamefont{Tsouris}},
  \bibinfo{journal}{J. Colloid Interface Sci.} \textbf{\bibinfo{volume}{206}},
  \bibinfo{pages}{532} (\bibinfo{year}{1998}).

\bibitem[{\citenamefont{Luo and Nguyen}(2017)}]{review_magnetic_flocculation}
\bibinfo{author}{\bibfnamefont{L.}~\bibnamefont{Luo}} \bibnamefont{and}
  \bibinfo{author}{\bibfnamefont{A.~V.} \bibnamefont{Nguyen}},
  \bibinfo{journal}{Sep. Purif. Technol.} \textbf{\bibinfo{volume}{172}},
  \bibinfo{pages}{85} (\bibinfo{year}{2017}).

\bibitem[{\citenamefont{Israelachvili}(1991)}]{Israelachvili}
\bibinfo{author}{\bibfnamefont{J.}~\bibnamefont{Israelachvili}},
  \emph{\bibinfo{title}{Intermolecular and Surface Forces}}
  (\bibinfo{publisher}{Academic Press, London}, \bibinfo{year}{1991}).

\end{thebibliography}

\end{document}